\documentstyle[emulateapj,graphics]{article}
\slugcomment{To appear in The Publications of the Astronomical Society of the Pacific, March 2001}

\newcommand{\kms}{$~{\rm km~s^{\scriptscriptstyle -1}}$}
\newcommand{\kmsMpc}{$~{\rm km~s^{\scriptscriptstyle -1}~Mpc^{\scriptscriptstyle -1}}$}
\newcommand{\etal}{et~al.} 
\newcommand{\eg}{e.g.}
\newcommand{\ie}{i.e.}
\newcommand{\delm}{$\Delta m_{15}$}
\newcommand{\lam}{$\lambda$}
\newcommand{\leib}{L93}  
\newcommand{\lleib}{L93}
\newcommand{\flip}{F92}
\newcommand{\fflip}{F92}
\newcommand{\tur}{T96}
\newcommand{\ttur}{T96}
\newcommand{\maz}{M97}
\newcommand{\mmaz}{M97}
\newcommand{\nick}{$^{56}$Ni}
\newcommand{\sm}{M$_{\odot}$}
\newcommand{\rrie}{Riess \etal~(1999b)}
\newcommand{\rie}{Riess \etal~1999b}

\begin{document}

\title{The Subluminous Type Ia Supernova 1998de in NGC 252 }
\author{Maryam Modjaz\altaffilmark{1}, Weidong Li\altaffilmark{2}, Alexei V. Filippenko\altaffilmark{2}, Jennifer Y. King\altaffilmark{1}, Douglas C. Leonard\altaffilmark{2}, Thomas Matheson\altaffilmark{3}, Richard R. Treffers\altaffilmark{2},}
\affil{Department of Astronomy, University of California, Berkeley, CA 94720-3411}

\author{and}

\author{Adam G. Riess\altaffilmark{4}} 
\affil{Space Telescope Science Institute, 3700 San Martin Drive, Baltimore, MD 21218}

\altaffiltext{1}{modjaz, jyking@ugastro.berkeley.edu}
\altaffiltext{2}{wli, alex, dleonard, rtreffers@astro.berkeley.edu}
\altaffiltext{3}{Present address: Center for Astrophysics, 60 Garden Street, Cambridge, MA 02138, tmatheso@cfa.harvard.edu}
\altaffiltext{4}{ariess@stsci.edu}

\begin{abstract}
	We present spectroscopic and extensive photometric observations of supernova (SN) 1998de in the S0 galaxy NGC 252, discovered during the course of the Lick Observatory Supernova Search. These data, which span a time period of 8 days before to 76 days after $B$-band maximum, unambiguously establish SN~1998de as a peculiar and subluminous type Ia supernova (SN Ia) with strong similarities to SN~1991bg, the prototype of these intrinsically dim SNe Ia. We find that SN~1998de, which has \delm($B$)=1.95$\pm$0.09 mag, rises and declines much faster than normal SNe Ia and does not exhibit the usual plateau in the $R$ band. In the $I$ band it shows a short plateau phase or possibly a secondary maximum, soon after the first maximum. We find that subluminous SNe Ia with the same value of \delm($B$) can have slightly different light curves at longer wavelengths. The notable spectroscopic similarities between SN~1998de and SN~1991bg are the wide \ion{Ti}{2} trough at 4100$-$4500~\AA, the strong \ion{Ca}{2} features, and the early onset of the nebular phase. We observe that spectroscopic deviations of SN~1998de from SN~1991bg increase toward redder wavelengths. These deviations include the absence of the conspicuous \ion{Na}{1}~D absorption found in SN~1991bg at 5700~\AA, and the evolution of a region (6800$-$7600~\AA) from featureless to feature-rich. Several lines of evidence suggest that SN~1998de was a slightly more powerful explosion than SN~1991bg. We discuss the implications of our observations for progenitor models and the explosion mechanism of peculiar, subluminous SNe Ia. The extensive photometric data make SN~1998de a better template than SN~1991bg for calibrating the low-luminosity end of the luminosity vs. decline-rate relationship.
\end{abstract} 


\section{INTRODUCTION}
	Supernovae (SNe) are classified into two main categories according to whether there are hydrogen lines present in their spectrum, usually taken near peak brightness: hydrogen-deficient supernovae are designated as Type I (SNe I), while those with hydrogen are Type II (SNe II). The category of SNe I is again subdivided into SNe Ia, Ib, or Ic, depending on whether they show further spectral features. Type Ia supernovae (SNe Ia) are identified by the appearance of a strong absorption feature at about 6150~\AA~due to blueshifted \ion{Si}{2} \lam 6355 (for reviews see, \eg, Filippenko 1997; Harkness \& Wheeler 1990). 
	
	A striking feature of SNe Ia has been their homogeneity; they have only moderate scatter in observed absolute peak magnitudes (Hamuy \etal~1996a) and similar light-curve shapes and spectra (\eg, Leibundgut \& Pinto 1992; Branch \& Tamman 1992; Filippenko 1997; Branch 1998). Moreover, the observed variations in luminosity among SNe Ia are found to be correlated with their light-curve shape, hence furnishing a tool to accurately calibrate the peak luminosity of SNe Ia (\eg, the \delm($B$)~method --- Phillips 1993; Hamuy \etal~1996a,b; Phillips \etal~1999; the multicolor light-curve shape method --- Riess, Press, \& Kirshner 1996; Riess \etal~1998a; the stretch method --- Perlmutter \etal~1997; the two-parameter method --- Tripp 1998). This has led to the successful usage of SNe Ia as ``calibrated candles'' for both nearby and cosmological distances, and in the context of an array of applications such as tracing the expansion history of the Universe (Riess \etal~1998a; Perlmutter \etal~1999), measuring the Hubble constant (\eg, Branch 1998; Suntzeff \etal~1999; Jha \etal~1999), studying the peculiar motions and bulk flows of galaxies (\eg, Riess, Press, \& Kirshner 1995; Zehavi \etal~1998; Riess 2000), and probing the chemical evolution of galaxies (see, \eg, Pagel 1997 for a review).

	In 1991, however, two nearby objects showed that there can occasionally be extreme deviants among SNe Ia: SN~1991bg, an intrinsically very dim SN Ia (Filippenko \etal~1992b, hereafter referred to as F92; Leibundgut \etal~1993, hereafter L93; Ruiz-Lapuente \etal~1993; Turatto \etal~1996, hereafter T96; Mazzali \etal~1997, hereafter M97), and SN~1991T (Filippenko \etal~1992a; Ruiz-Lapuente \etal~1992; Phillips \etal~1992; Jeffery \etal~1992), an intrinsically overluminous object.\footnote{Note that the new Cepheid distance to NGC~4527, the host galaxy of SN~1991T, derived by Saha, Labhardt, \& Posser (2000) suggests that SN~1991T is not as overluminous as previously claimed.} Both of these SNe showed clear spectral differences from typical SNe Ia. Over the last decade, more instances have been found which exhibit distinct photometric and spectroscopic peculiarities, and it is now commonly thought that this group of so-called ``peculiar'' SNe Ia can be subdivided into SN~1991bg-like (subluminous) objects and SN~1991T-like (overluminous) objects. Although Branch, Fisher, \& Nugent (1993) found that approximately 85\% of all previously observed SNe Ia were normal (``Branch-normal''), recent results by Li \etal~(2000b; 2001a,b) suggest an intrinsic peculiarity rate as high as 36\% (20\% SN~1991T-like, 16\% SN~1991bg-like).

	Whereas many normal SNe Ia have been studied in detail and well-sampled light curves have been published (\eg, Riess \etal~1999a), only a few peculiar SNe Ia (specifically of the subluminous variety) have been thoroughly observed. Examples include SN~1991bg, SN~1992K (Hamuy \etal~1994), and SN~1997cn (Turatto \etal~1998). A detailed study is necessary to address concerns about the utility of peculiar SNe Ia as reliable distance indicators, as well as to answer questions about their progenitor systems and explosion mechanisms.
 
	In this paper we present SN~1998de as another instance of a peculiar, subluminous SN Ia. It was the eighth SN discovered by the Lick Observatory Supernova Search (LOSS; Treffers \etal~1997; Li \etal~2000a; Filippenko \etal~2001) with the 0.75-m Katzman Automatic Imaging Telescope (KAIT; Richmond, Treffers, \& Filippenko 1993; Filippenko \etal~2001). A description of our observations and analysis, including methods of performing photometry, calibration of the measurements onto the standard Johnson-Cousins system, the resulting multicolor light curves, and comparisons between the light curves of SN~1998de, SN~1991bg, and other SNe Ia is given in \S2. The absolute magnitudes of SN~1998de are presented in \S3. Section 4 contains a description of the observations and analysis of the spectra of SN~1998de, including a thorough comparison of the spectra of SN~1998de, SN~1991bg, and the normal SN Ia 1994D. We discuss the implications of our observations on constraining progenitor models in \S5 and summarize our conclusions in \S6. 

\section{PHOTOMETRY}

\subsection{Observations and Reductions}

	SN~1998de, located 72\arcsec~from the nucleus of the S0 galaxy NGC 252, was discovered at $m \approx$ 18.4 mag by Modjaz \etal~(1998) in an unfiltered image obtained on 1998 July 23.5 and confirmed on July 24.5 during the LOSS.\footnote{We use UT dates throughout this paper.} A KAIT image on July 19.5 already revealed a hint of the supernova, whereas a KAIT image on July 14.5 with a limiting magnitude of about 19 did not show any star at the supernova's position, $\alpha=0^h 48^m 06^s.88$ and $\delta=+27\arcdeg 37\arcmin 28{\farcs}5$ (equinox J2000.0). Due to the search's small baseline (that is, the time interval to repeat observations of target galaxies; $\sim$ 3 to 5 days), this supernova was caught 8 days prior to maximum brightness (on 1998 July 31.8 in $B$, see \S2.2). Garnavich, Jha, \& Kirshner (1998) classified a spectrum of SN~1998de obtained on July 25.4 as Type Ia by the presence of a broad absorption feature of \ion{Si}{2} at 6185 \AA, noting that \ion{Ti}{2} absorption at the blue end of the spectrum and strong \ion{Si}{2} at 5800~\AA~suggested a peculiar and subluminous event, similar to SN~1991bg.
	
	A KAIT monitoring program of broadband $BVRI$ photometry was established for SN~1998de two days after discovery and it continued for about 2 months. Also, optical spectra of SN~1998de were obtained sporadically with the 3-m Shane reflector at Lick Observatory 1$-$2.5 months past maximum brightness. The photometric data were collected with the 512 $\times$ 512 pixel CCD camera (AP7) made by Apogee Instruments at the $f$/8.2 Cassegrain focus of KAIT with exposure times in the range of 300 s (in $R$ and $I$) and 600 s (in $B$ and $V$). The scale was 0{\farcs}8~per pixel and the field of view was 6{\farcm}7~$\times$~6{\farcm}7. The typical seeing at Lick was around 3\arcsec~full width at half maximum (FWHM). The bias and dark-current subtraction with subsequent twilight-sky flatfielding were accomplished automatically at the telescope. Significant fringing in the $I$ band could not be completely removed, but this phenomenon has only a small impact on our results because the SN and comparison stars were considerably brighter than the background. This problem is presumably due to the fact that the infrared sky lines vary in brightness over time.

	The telescope's autoguiding device failed between July 25 and July 28. Multiple 60-s exposures were taken in each filter during that time, since the tracking mechanism was sufficiently accurate for exposure times of ~$\la$100 s. They were later aligned and combined to form a single frame in each filter.

\vspace{0.09in}
{\plotfiddle{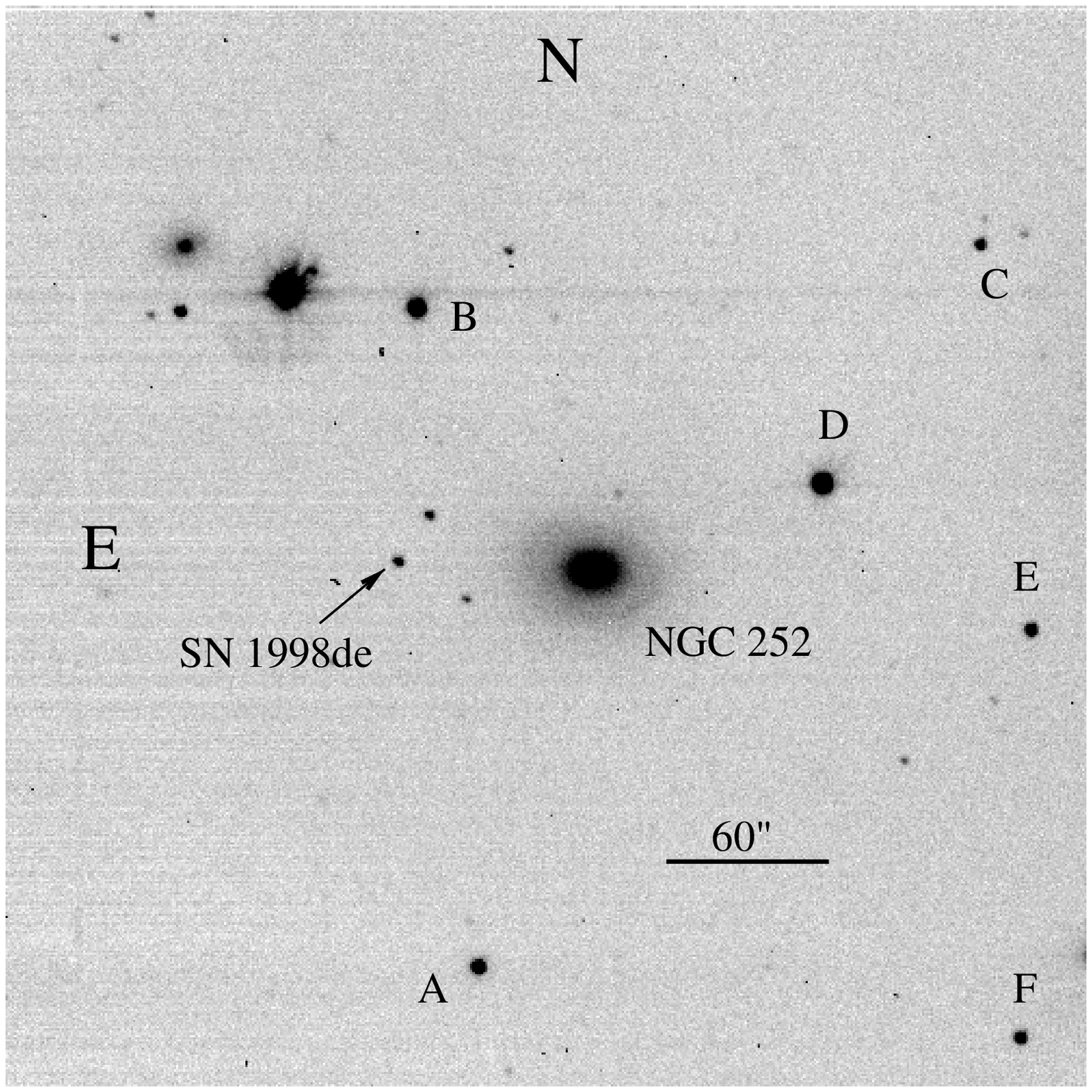}{3.2in}{0}{50}{50}{-30}{-70}}
{\footnotesize FIG. 1. --- \emph{V}-band KAIT image of the field of SN~1998de, taken on 1998 July 31. The field of view is 6{\farcm}7 by 6{\farcm}7. The six local standard stars are marked (A$-$F), and SN~1998de is about 72\arcsec~east and 3\arcsec~north of the nucleus of NGC~252.

}
\vspace{0.08in}

\begin{deluxetable}{ccccccccc}
\tablecaption{Magnitudes of Local Standard Stars}
\tablehead{
\colhead{Star} & 
\colhead{$B$} & \colhead{$\sigma_B$} & \colhead{$V$} &
\colhead{$\sigma_V$} & \colhead{$R$} & \colhead{$\sigma_R$} & \colhead{$I$}
& \colhead{$\sigma_I$}  
}
\startdata
     A   & 15.701 & 0.005 &14.860 &0.003 &14.390 &0.001 &13.996 &0.007 \\
     B   & 14.462 & 0.009 &13.708 &0.003 &13.268 &0.003 &12.916 &0.003 \\
     C   & 17.460 & 0.010 &16.541 &0.004 &16.012 &0.007 &15.520 &0.011 \\
     D   & 13.736 & 0.005 &13.092 &0.003 &12.733 &0.003 &12.443 &0.003 \\ 
     E   & 16.346 & 0.003 &15.652 &0.004 &15.267 &0.008 &14.971 &0.006  \\
     F   & 16.486 & 0.006 &15.727 &0.003 &15.315 &0.010 &14.988 &0.011 \\
\enddata
\end{deluxetable}

	Figure 1 shows a KAIT $V$-band image taken on 1998 July 31 with SN~1998de and six comparison stars marked. Absolute calibration of the field was done on 1998 September 19, 20, and 21 by observing Landolt (1992) standard stars at different airmasses throughout the photometric nights. Instrumental magnitudes for the standards were measured by employing simple aperture photometry with the IRAF\footnote{IRAF (Image Reduction and Analysis Facility) is distributed by the National Optical Astronomy Observatories, which are operated by the Association of Universities for Research in Astronomy, Inc., under cooperative agreement with the National Science Foundation.} DAOPHOT package (Stetson 1987) and then used to determine transformation coefficients to the standard systems of Johnson (Johnson \etal~1966, for \emph{BV}) and Cousins (Cousins 1981, for \emph{RI}). The derived transformation coefficients and color terms were used to calibrate the sequence of six stars near SN~1998de that functioned as comparison stars for the supernova's brightness. The magnitudes of these six local standard stars and the associated uncertainties derived by averaging over the three photometric nights are listed in Table 1. Since the field of view between July 25 and July 28 did not include the local standards E and F, they were not used when computing the brightness of SN~1998de for those dates.

\begin{deluxetable}{ccccccc}
\tablecaption{Photometric Observations of SN 1998de}
\tablehead{
\colhead{UT Date} & \colhead{JD $-$ } & \colhead{Phase\tablenotemark{a}} &
\colhead{$B$ ($\sigma_B$)} & \colhead{$V$ ($\sigma_V$)} & \colhead{$R$ ($\sigma_R$)} & \colhead{$I$ ($\sigma_I$)}\\
 DD/MM/YY & 2,451,000.0 & (days)
}
\startdata
23/07/98                  & 18.0  & $-8$  & \nodata      & 18.63\tablenotemark{b} (0.10) &\nodata       & \nodata \\ 
24/07/98		  & 19.0  & $-7$  & \nodata      & 18.27\tablenotemark{b} (0.08)  &\nodata       & \nodata  \\
25/07/98   		  & 20.0  & $-6$  &   \nodata    & 17.80 (0.05) & 17.57 (0.03) & 17.70 (0.05) \\
26/07/98   		  & 21.0  & $-5$  & 18.37 (0.06) & 17.56 (0.03) & 17.40 (0.03) & 17.42 (0.05)\\
27/07/98   		  & 22.0  & $-4$  &   \nodata    & 17.37 (0.03) & 17.12 (0.03) & 17.14 (0.07) \\
28/07/98   		  & 23.0  & $-3$  & 17.84 (0.04) & 17.26 (0.03) & 17.02 (0.02) & 17.05 (0.06)\\
29/07/98   		  & 24.0  & $-2$  & 17.73 (0.04) & 17.09 (0.03) & 16.92 (0.03) & 16.95 (0.04)\\
30/07/98    		  & 25.0  & $-1$  & 17.62 (0.04) & 16.96 (0.02) & 16.81 (0.03) & 16.83 (0.03)\\
31/07/98    		  & 26.0  & $0 $  & 17.55 (0.03) & 16.93 (0.03) & 16.76 (0.03) & 16.75 (0.04) \\
01/08/98   		  & 27.0  & $+1$  & 17.58 (0.04) & 16.85 (0.03) & 16.63 (0.03) & 16.67 (0.04) \\
02/08/98    		  & 28.0  & $+2$  & 17.60 (0.03) & 16.83 (0.02) & 16.61 (0.03) & 16.70 (0.03) \\
03/08/98    		  & 29.0  & $+3$  & 17.71 (0.03) & 16.81 (0.02) & 16.62 (0.02) & 16.62 (0.04)  \\
04/08/98  	  	  & 30.0  & $+4$  & 17.77 (0.03) & 16.82 (0.03) & 16.58 (0.02) & 16.53 (0.03)  \\
06/08/98    		  & 32.0  & $+6$  & 18.08 (0.04) & 16.98 (0.02) & 16.70 (0.03) & 16.67 (0.05) \\
07/08/98   		  & 33.0  & $+7$  & 18.29 (0.06) & 17.04 (0.02) & 16.74 (0.03) & 16.68 (0.04) \\
08/08/98  	          & 33.9  & $+8$  & 18.47 (0.08) & 17.13 (0.03) & 16.80 (0.02) & 16.66 (0.03)  \\
09/08/98   		  & 34.9  & $+9$  & 18.78 (0.08) & 17.23 (0.03) & 16.88 (0.02) & 16.66 (0.03)  \\
11/08/98\tablenotemark{c} & 37.0  & $+11$ &    \nodata   & 17.39 (0.06) & 16.98 (0.04) & 16.66 (0.08) \\
12/08/98\tablenotemark{c} & 38.0  & $+12$ & 19.19\tablenotemark{d} (0.12) & 17.51 (0.08) & 17.11 (0.07) & \nodata \\
13/08/98\tablenotemark{c} & 39.0  & $+13$ &    \nodata   & 17.74 (0.08) & 17.20 (0.05) &  \nodata  \\
15/08/98   		  & 41.0  & $+15$ & 19.46 (0.11) & 17.85 (0.04) & 17.36 (0.03) & 16.91 (0.04)\\
16/08/98  	          & 41.8  & $+16$ &    \nodata   & 17.96 (0.04) & 17.45 (0.03) & 16.99 (0.04)\\  
17/08/98  	          & 43.0  & $+17$ & 19.63\tablenotemark{d} (0.06) & 18.01 (0.04) & 17.56 (0.03) & 17.07 (0.04)      \\
18/08/98   		  & 44.0  & $+18$ &    \nodata   & 18.18 (0.04) & 17.68 (0.03) & 17.14 (0.04)   \\
19/08/98   		  & 44.8  & $+19$ & 19.63 (0.07) & 18.27 (0.04) & 17.78 (0.04) & 17.27 (0.05)     \\
21/08/98   		  & 46.8  & $+21$ & 19.83 (0.08) & 18.47 (0.05) & 17.97 (0.03) & 17.45 (0.05)     \\
23/08/98   		  & 48.9  & $+23$ & 19.99 (0.11) & 18.58 (0.06) & 18.14 (0.03) & 17.66 (0.04)    \\
25/08/98   		  & 50.8  & $+25$ & 19.95 (0.12) & 18.64 (0.05) & 18.31 (0.04) & 17.72 (0.06)    \\
27/08/98    		  & 52.8  & $+27$ & 20.11 (0.17) & 18.84 (0.06) & 18.41 (0.04) & 17.87 (0.05)    \\
29/08/98   		  & 54.8  & $+29$ & 20.20 (0.20) & 18.97 (0.06) & 18.57 (0.04) & 18.05 (0.06)  \\
31/08/98   		  & 56.8  & $+31$ & \nodata      & 19.02 (0.09) & 18.61 (0.05) & 18.08 (0.07)    \\
10/09/98   		  & 67.0  & $+41$ & \nodata      & 19.32 (0.19) & 18.98 (0.13) & 18.84 (0.20)    \\
15/09/98   		  & 71.8  & $+46$ & \nodata      & 19.44 (0.07) & 19.19 (0.05) & 19.04 (0.14)  \\
20/09/98\tablenotemark{d} & 76.8  & $+51$ & 20.83 (0.20) & 19.59 (0.08) & 19.49 (0.06) & 19.18 (0.09)   \\
\enddata
\tablenotetext{a}{Relative to the epoch of $B$ maximum (JD$=$2,451,026.3).}
\tablenotetext{b}{Originally ``clear'' observation.}
\tablenotetext{c}{Bright full moon.}
\tablenotetext{d}{Measured from combined images.}
\end{deluxetable}

	We implemented the point-spread-function (PSF) fitting technique for the differential photometry of the supernova relative to the comparison stars. Due to the fortunate position of the supernova far from the nucleus of the host galaxy in a region with a faint and smoothly varying background, measuring the brightness of the supernova was not hampered by a complex environment, which usually necessitates template subtraction (\eg, Filippenko \etal~1986; Richmond \etal~1995). We could have chosen simple aperture photometry as our technique, but we decided against it; besides giving more accurate results than aperture photometry for objects projected upon slightly more complex regions (Schmidt \etal~1993), PSF fitting has the advantage of creating an image in which the measured objects can be subtracted. This can serve as an indicator of the quality of the brightness measurements of the comparison stars and the supernova. 

	Before the photometry procedure itself was initiated, special care was taken to remove cosmic rays, especially those around the PSF stars. Since our particular CCD camera is cooled thermoelectrically rather than with liquid nitrogen (for mechanical simplicity, and to decrease human maintenance of the robotic telescope), the temperature of the camera is not very stable and dark current is not always cleanly removed from the images. The resulting low-level dark-current residual varies during the night and has to be removed manually by adding or subtracting a few percent of the long-exposure dark-current image. The uncertainties introduced by the manual scaling of the dark current have only negligible effects on the brightness measurement of the high-contrast SN and the PSF stars. 

	The PSF for each image was determined with DAOPHOT using well-isolated stars in the field, which are neither too bright (saturated) nor too faint. We used the inner core of SN~1998de and the local standards to fit the PSF in order to increase the signal-to-noise (S/N) ratio. To measure the sky backgrounds of the SN and the local standards, we used an annulus with an inner radius of always 20 pixels (16\arcsec) and an outer radius of 25-28 pixels (20\arcsec-22\arcsec), depending on the seeing. 

	The instrumental magnitudes measured from the PSF-fitting method were transformed onto the Johnson-Cousins system. Employing the color terms derived from calibration observations on 1998 September 19, 20, and 21, the linear transformation equations used were of the following form:

\noindent \begin{equation}    b = B - 0.086 (B - V) + C_B,\end{equation}
\begin{equation}    v = V + 0.029 (B - V) + C_V,\end{equation}
\begin{equation}    r = R - 0.151 (V - R) + C_R,\end{equation}        	
\begin{equation}    i = I - 0.057 (V - I) + C_I.\end{equation}    
    		
\noindent In the equations above, the upper-case passband letters signify magnitudes in the standard Johnson-Cousins system, whereas the lower-case letters denote instrumental magnitudes. The constants $C_B, C_V, C_R$, and $C_I$ are the differences between the zero-points of the instrumental and standard magnitudes.\footnote{Our Johnson $R$ filter, which gave rise to the rather large color term, has now been replaced by a Cousins/Bessell one, completing our new set of $UBVRI$ filters.} Atmospheric extinction corrections were not considered since the net effect of this extinction is expressed through an additive constant that vanishes when doing differential photometry, i.e., when measuring the brightness of SN~1998de relative to the comparison stars.  

	Since the search itself is conducted using unfiltered CCD observations for efficiency reasons (an unfiltered CCD image can reach fainter stars in less time than a filtered image), the magnitudes from the discovery and confirmation nights had to be transformed to a standard passband in order to compare with subsequent filtered observations. The employed method, which has also been used by \rrie, treats the unfiltered data as observations in the pseudo-passband ``Clear'' ($C$), so that the conversion equation into the standard passband $V$ is simply expressed by the pseudo-color $V - C$. We chose to transform the unfiltered data to the standard passband $V$, since the $V$ response curve is the closest match to the unfiltered response function. The results of detailed examinations by \rrie~of the involved conversion parameters (\eg, the response function of KAIT's CCD) yield a linear relation for stars between the standard color $B - V$ and the pseudo-color of the following form:

 \noindent \begin{equation} V - C = C_{VC} (B - V). \end{equation}

\noindent Here $V$ is the standard Johnson passband $V$, and $C_{VC}$ is the slope of the linear relationship between the standard color \bv~and the pseudo-color $V - C$, empirically measured to be 0.33 $\pm$ 0.10 for the KAIT passbands (\rie) having imposed the condition that $V - C = 0$ when $B - V = 0$.

	Although the spectra of supernovae do not resemble those of normal stars (\eg, Filippenko 1997), their photometric behavior at very early times can be best approximated with a thermal model; their spectra are dominated by a thermal continuum (see \rie~for a detailed discussion). Thus, with the instrumental magnitudes for the unfiltered observations, the transformation equations for the supernova (SN) and the local standards (LS) are 
 
\noindent \begin{equation} V_{SN} - c_{SN} = C_{VC} (B - V)_{SN} + A,  \end{equation} 
\begin{equation} V_{LS} - c_{LS} = C_{VC} (B - V)_{LS} + A.	\end{equation}

\noindent Here the lower-case letter $c$ denotes the instrumental magnitudes in the pseudo-passband ``Clear'', and $A$ an airmass extinction term that is the same for both SN and local standards in the same frame. Combining the two equations we can infer the standard $V$ magnitude of the supernova,

\noindent \begin{equation} V_{SN} = V_{LS} + (c_{SN} - c_{LS}) + C_{VC} [(B - V)_{SN} - (B - V)_{LS}].\end{equation}

	 The amount of color evolution of the infant supernova, encapsulated by the term $(B - V)_{SN}$ for the discovery and confirmation observations, has a direct impact on our derived values for the standard magnitudes and cannot be directly measured, but rather has to be derived from later points. We chose the value of $(B - V)_{SN}$ on the dates of the unfiltered observations (8 and 7 days before maximum brightness in the $B$ passband) to be the same as on $t$ = $-$5 days, since the color of SN~1998de seems to evolve slowly at early times, as indicated by our data and in accordance with the behavior of most SNe Ia.

	Despite the relative proximity of SN~1998de ($cz =$ 4950\kms; see \S3), we applied $K$ corrections in order to account for the redshift effect. These are larger for SN~1998de than for normal SNe Ia, presumably due to its red color (see \S2.3; as Hamuy \etal~1993 show, there seems to be a tight relationship between color evolution and $K$-term curve). The $K$ corrections in $B,V$, and $R$ were calculated using the spectra of SNe~1991bg and~1998de, but since most of the available spectra do not encompass the $I$-band wavelengths, the corrections in $I$ are taken from normal SNe Ia.

	Table 2 shows our final results for all photometric observations ranging from 8 days prior to $B$ maximum until 51 days after maximum brightness. Uncertainties were estimated by combining in quadrature the errors given by the error analysis package in DAOPHOT with those introduced by the transformation of instrumental magnitudes onto the standard system and by the $K$ correction. Since SN~1998de was a relatively faint and red supernova, bad weather conditions influenced our data and affected the $B$ images the most. For the sets of nights of August 11/12/13, August 16/17/18, and September 19/20/21, the $B$ images had to be combined to yield one single image for each set in order to obtain a sufficiently high S/N ratio. The data for September 19/20/21, the last three nights of observation, were combined in this manner for all filters, since the SN was already very faint at that point. For the combined images, the epoch of the observation was taken to be August 12, August 17, and September 20, respectively.

\subsection{Optical Light Curves}

\vspace{0.45in}
{\plotfiddle{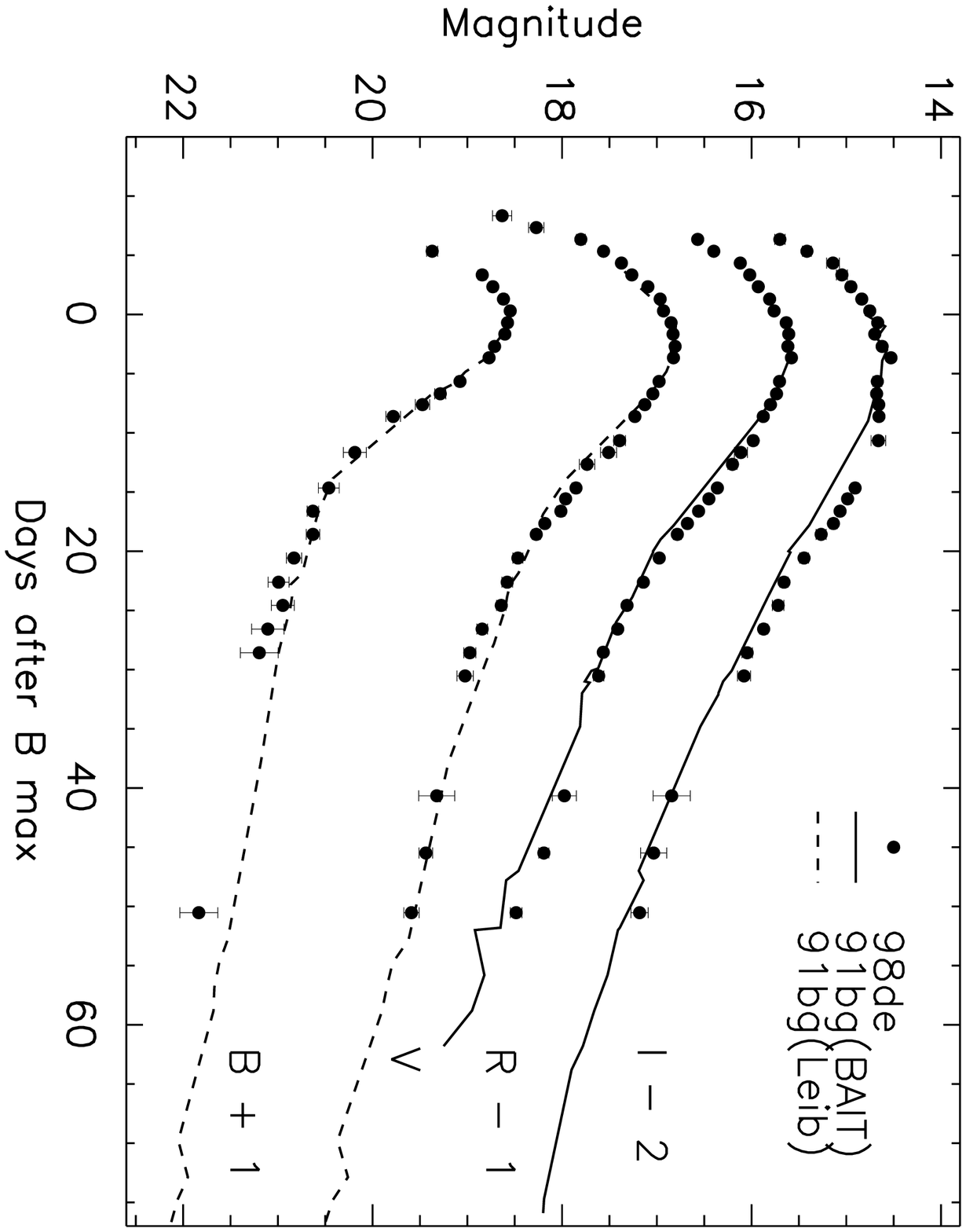}{3.2in}{90}{45}{48}{290}{-10}}
{\footnotesize FIG. 2. --- The $B, V, R,$ and $I$ light curves of SN~1998de, along with those of SN~1991bg ($RI$ from \flip,``BAIT''; $BV$ from \leib,``Leib''), which are shifted in time and peak magnitude to match those of SN~1998de. The time zero-point is the date of maximum light in the $B$ passband, here 1998 July 31.8 UT.

}
\vspace{0.08in}


	Figure 2 displays our $BVRI$ light curves of SN~1998de together with those of SN~1991bg (\flip; \leib). For all SN~1998de points except the marked ones, the uncertainties are smaller than the plotted symbols. The values for the date and the magnitude of the peak in each filter are given in Table 3; they were determined by fitting different spline functions and polynomials to the $BVRI$ light curves around maximum brightness. Since we have ample pre- and post-maximum data points, the uncertainties in our values are comparatively small, except in the $I$ band ($\sigma_I=$1.1 d), where there does not seem to be one clear peak. Here, and in all subsequent discussions and figures, the variable $t$ denotes time since maximum brightness in the $B$ passband (JD=2,451,026.3).

\begin{deluxetable}{ccccc}
\tablecaption{Photometric Information on SN~1998de}   
\tablehead{
\colhead{Filter} & \colhead{$B$} & \colhead{$V$} &
\colhead{$R$} & \colhead{$I$} 
}
\startdata
Epoch of max.       & Jul. 31.8 $\pm$ 0.1 &   Aug. 3.0 $\pm$ 0.2  &  Aug. 3.6 $\pm$ 0.5 & Aug. 5.0 $\pm$ 1.1 \\
Julian Date of max. &      26.3 $\pm$ 0.1        & 28.5 $\pm$ 0.2      & 29.1 $\pm$ 0.5      & 30.5 $\pm$ 1.1 \\
  $-$ 2,451,000.0 &&&& \\
Magnitude at max.   & 17.56 $\pm$ 0.04    & 16.82 $\pm$ 0.04    & 16.61 $\pm$ 0.04    & 16.61 $\pm$ 0.05\\
$\Delta m_{15}$     & 1.95 $\pm$ 0.09     & 1.31 $\pm$ 0.06     & 1.08 $\pm$ 0.06     & 0.70 $\pm$ 0.09
\enddata
\end{deluxetable}

	In Figures 3 and 4, we compare the light curves of SN~1998de with those of several other well-observed SNe Ia representing the diversity of SN Ia light curves: SN~1991T (Ford \etal~1993 for $R$; Hamuy \etal~1996b for $BVI$) as an example of an overluminous SN Ia; SN~1989B (Wells et al. 1994) and SN 1994D (Richmond \etal~1995) as examples of normal SNe Ia, and SN~1991bg (\flip~for $RI$; \leib~for $BV$) as the prototypical subluminous SN Ia. All light curves are shifted in time and peak magnitude to match those of SN~1998de with the time zero-point being the date of maximum light in the $B$ passband. The data for the SN~1991bg-like objects SN~1992K (Hamuy \etal~1994) and SN~1997cn (Turatto \etal~1998) are not explicitly included here, since they are claimed to be twins to SN~1991bg both photometrically and spectroscopically, but comments and comparisons are provided. 

	Figures 2$-$4 show unambiguously that the early-time light curves of SN~1998de closely resemble those of SN~1991bg. Nevertheless, there seems to be a certain trend, where the discrepancies may increase with increasing wavelength of the passband so that the closest match between the two SNe is in the $B$ passband (except at late times). On the other hand, the most notable similarities in all filters are the narrow peak, beautifully exhibited by SN~1998de, and the rapid decline rate distinguishing these two SNe from the rest of the sample. Here SN~1998de displays among the best premaximum observations in our chosen sample --- only SN~1994D was caught at a comparably early time. Unlike the case for SN~1991bg, for which there were no premaximum observations (or only a few in $V$; see \leib), \emph{our data clearly establish SN~1998de as a fast riser}.

	The $B$-band light curve of SN~1998de (Figure 3) matches that of SN~1991bg very closely at early times, but seems to decline faster at late times (for $t\ga$20 days). In accordance with Phillips (1993) and Hamuy et al. (1996a,b), we define $\Delta m_{15}(X$) to be the decline (in magnitudes) during in the first 15 days after maximum brightness in the passband $X$, and we measure $\Delta m_{15}(B)$=1.95$\pm$0.09 for SN~1998de. The uncertainty is the sum in quadrature of the uncertainty in the $B$-band peak magnitude, the change in the interpolated magnitude 15 days after peak (if the date of maximum is changed by the amount listed in Table 3), and the uncertainty of the observed magnitudes. This \delm~value is nearly identical to that of SN~1991bg [$\Delta m_{15}$($B$)=1.93, Hamuy \etal~1996b; $\Delta m_{15}$($B$)=1.95, \tur], and constitutes the largest value in our comparison sample of SNe Ia [$\Delta m_{15}$($B$)=1.31 for SN 1994D, Richmond \etal~1995; 0.94 for SN 1991T, Hamuy \etal~1996b; 1.31 for SN 1989B, Wells \etal~1994] and of all SNe Ia sampled during the Cal\'{a}n/Tololo Supernova Survey (Hamuy \etal~1996a).

\vspace{0.30in}
{\plotfiddle{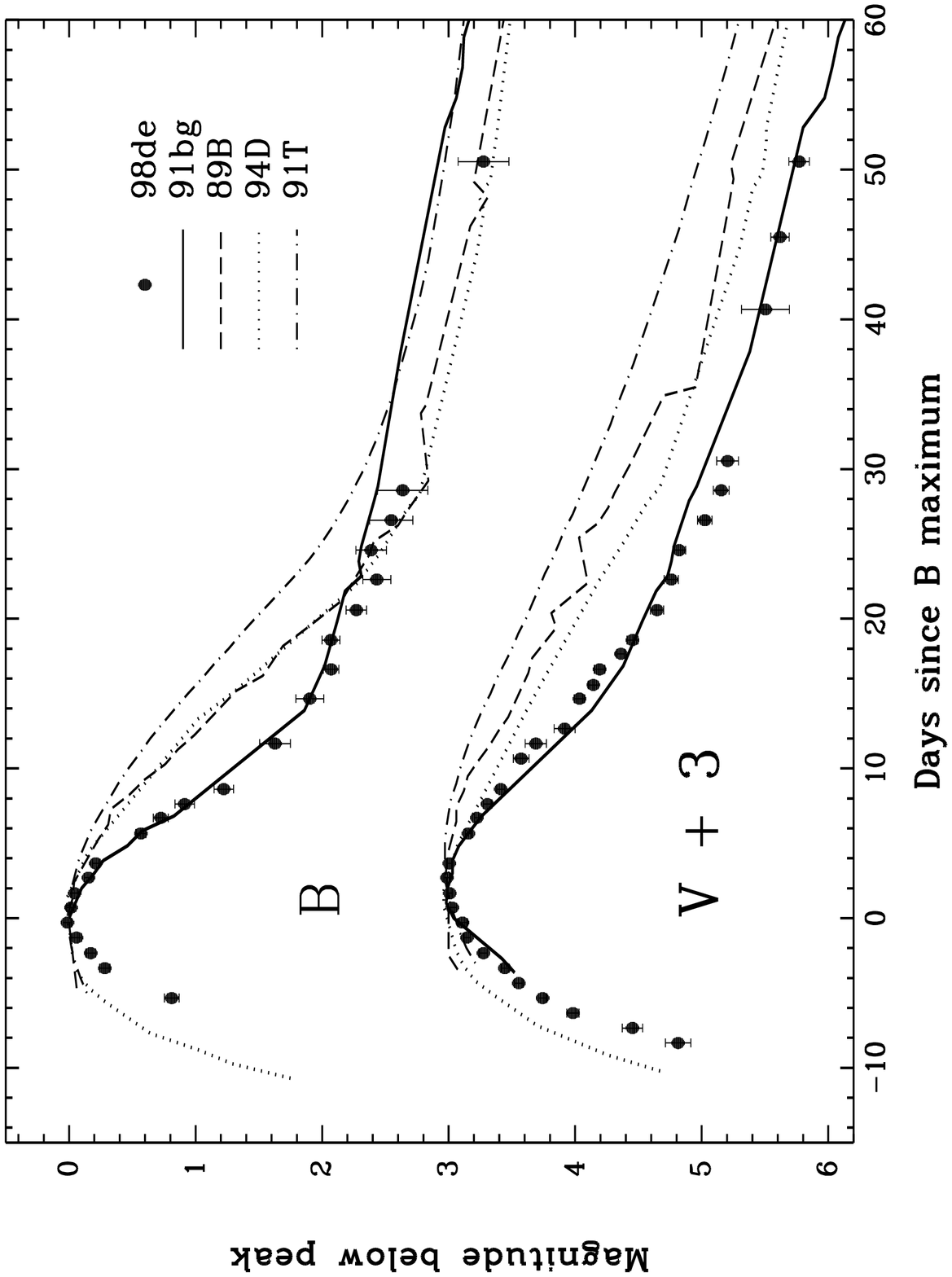}{3.2in}{-90}{40}{45}{-30}{270}}
{\footnotesize FIG. 3. --- The $B$ and $V$ light curves of SN~1998de, together with those of SN~1991bg(\leib), SN~1991T (Hamuy \etal~1996b), SN~1989B (Wells \etal~1994), and SN~1994D (Richmond \etal~1995). All light curves are shifted in time and peak magnitude to match those of SN~1998de. 

}
\vspace{0.12in}

	There seems to be a steadily growing deviation from the SN~1991bg curve at late times (for $t\ga$20 days), although the uncertainties of each individual measurement are relatively large. We can believe this deviation, since all data points of SN~1998de lie consistently below the SN~1991bg curve, and we conclude that it is due to an intrinsic difference between the photometric behavior of the two SNe at late times.
	
	The $V$-band curve is shown in Figure 3, and reveals consistent similarities between SN~1998de and SN~1991bg, reinforcing the narrow-peak behavior. We measure $\Delta m_{15}(V)$=1.31$\pm$0.06, slightly smaller than the corresponding value for SN~1991bg, $\Delta m_{15}(V)$=1.42 (\leib). 

	We present the $R$-band light curve in Figure 4 and note in SN~1998de the absence of the period of slower decline, the so-called ``plateau,'' which is characteristic of all normal SNe Ia at $t \approx$~20 to 30 days (Ford et al. 1993; Wells et al. 1994; Riess \etal~1999a). Rather, SN~1998de fades monotonically in $R$ (with \delm$(R)=1.08\pm0.06$), a bit slower than SN~1991bg, (\delm($R)=1.18$, based on our analysis of the \flip~data). There seems to be a region around the $R$-band maximum of SN~1998de that is not smooth, but with the present data set we cannot assess the reality of this. Perhaps significantly, the apparent ``wiggle'' appears to be even more pronounced in the $I$ band (Figure 4). Although these data points may say something about the supernova physics, we rather believe that they are two uncorrelated statistical deviations in the same direction.

\vspace{0.30in}
{\plotfiddle{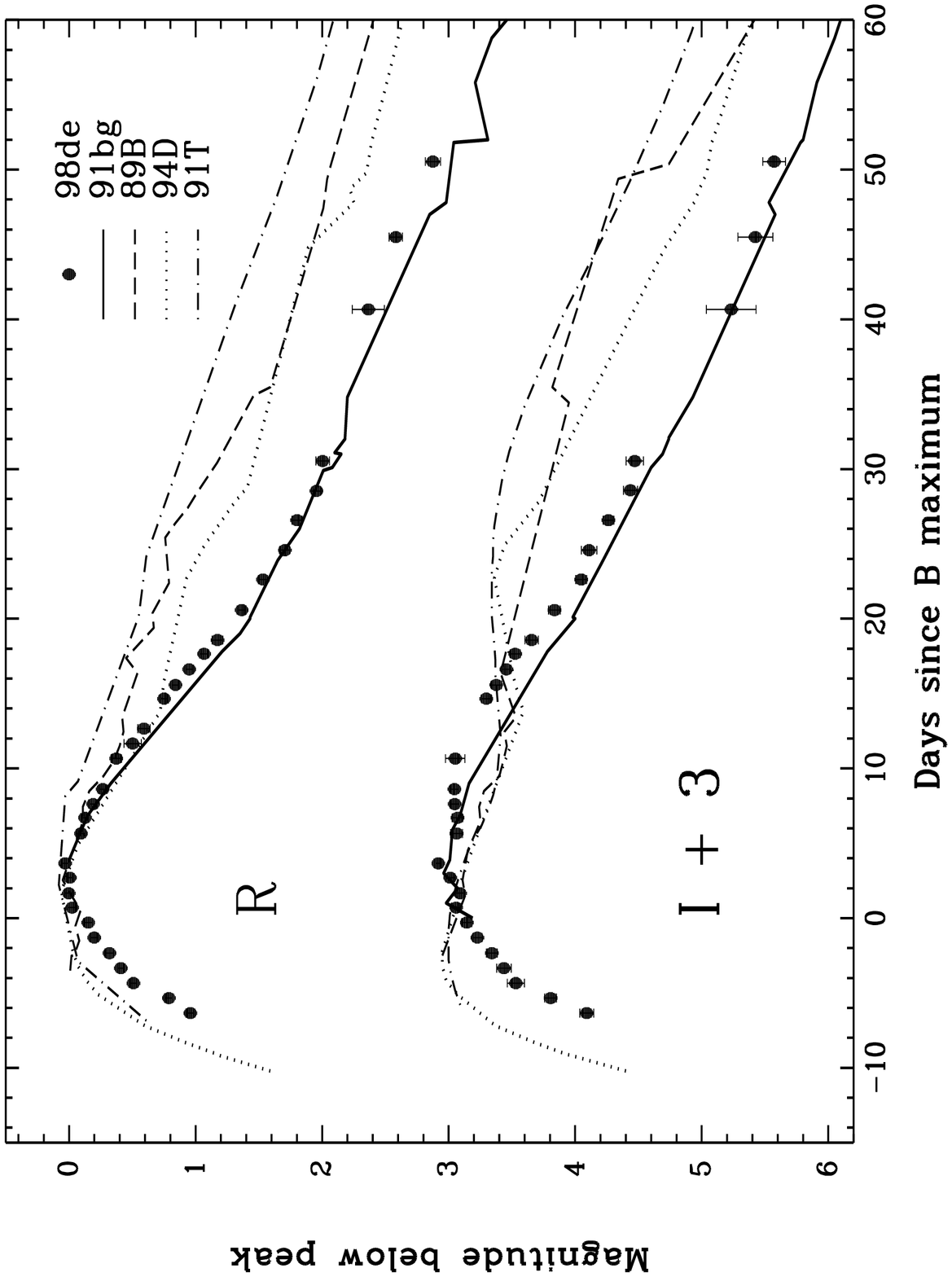}{3.2in}{-90}{40}{45}{-30}{270}}
{\footnotesize FIG. 4. --- Same as Fig. 3 but for the \emph{R} and $I$ light curves. The \emph{R} light curve of SN~1991T comes from Ford \etal~(1993). The \emph{R} and \emph{I} light curves of SN~1991bg come from \fflip.  

}
\vspace{0.12in}

	After the initial peak in the $I$ band 3$-$4 days past $B$ maximum (Figure 4), there is a plateau that extends out to $t \approx$~11 days. However, the $I$ band does not show the kind of second maximum whose prominence is usually expected of normal SNe Ia, again very similar to the $I$-band behavior of SN~1991bg (though the plateau in SN~1998de is more pronounced than that of SN~1991bg). The two other well-studied SN~1991bg-like events, SN 1992K (Hamuy et al. 1994) and SN1997cn (Turatto et al. 1998), were observed only well after maximum light, so that there is no range of overlap. More comparisons with new SN~1991bg-like objects are needed to see whether one maximum, two maxima, or a maximum plus plateau are typically present in these kinds of SNe Ia.

	It is interesting to note that if the $I$-band light curve is treated as having one broad peak, then its maximum in $I$ is reached 8 or 9 days later in SN~1991bg-like SNe Ia when compared with normal SNe Ia, falling between the two maxima exhibited by the latter. The derived \delm$(I)$~value is 0.70$\pm$0.09 for SN~1998de and 0.88 for SN~1991bg (based on our analysis of the \flip~data).

	We can also observe another trend, whose effect increases toward longer (redder) filter wavelength: although SN~1998de seems to concur completely with SN~1991bg in the $B$ filter at early times, the deviations become stronger in the other filters, with the post-maximum light curves declining more slowly than those of SN~1991bg. This trend is embodied in the smaller $\Delta m_{15}$ in all passbands (except $B$; see Table 3) when compared with those of SN~1991bg. 
   
	In summary, then, the light curves of SN~1998de closely resemble those of SN~1991bg, and deviate notably from those of normal SNe Ia such as SN~1994D and overluminous ones like SN~1991T. However, there are apparent differences between the light curves of SNe 1991bg and 1998de, especially at redder wavelengths, so we cannot simply regard SN~1998de as a mere twin of SN~1991bg. This conclusion, which is based on early- and intermediate-time photometry of SN~1998de, will be strengthened by the intermediate- and late-time spectral analysis in \S4. 

\subsection{Optical Color Curves}

	In Figure 5 we present the intrinsic optical color curves of SN~1998de, $(B - V)_0$, $(V - R)_0$, and $(V - I)_0$, together with color curves of several other SNe Ia (1991bg, 1989B, 1994D, and 1991T) for comparison. For SN~1998de, a Galactic reddening (giving rise to a color excess of $E(B-V)$=0.06 mag) was derived using full-sky maps of dust infrared emission (Schlegel, Finkbeiner, \& Davis 1998). In order to calculate the color excess for the $I$ and $R$ passbands we adopt the standard reddening law parameterization, $E(V-I)/E(B-V)=1.60$ and $E(V-R)/E(B-V) =0.78$ (Savage $\&$ Mathis 1979). We follow Wells \etal~(1994) in correcting the data for SN 1989B with $E(B-V)$=0.37 mag, and we assume reddening $E(B-V)$ of 0.04 mag for SN1994D (Richmond et al. 1995), 0.13 mag for SN~1991T (Phillips \etal~1992), and 0 mag for SN~1991bg (\flip).

\vspace{0.25in}
{\plotfiddle{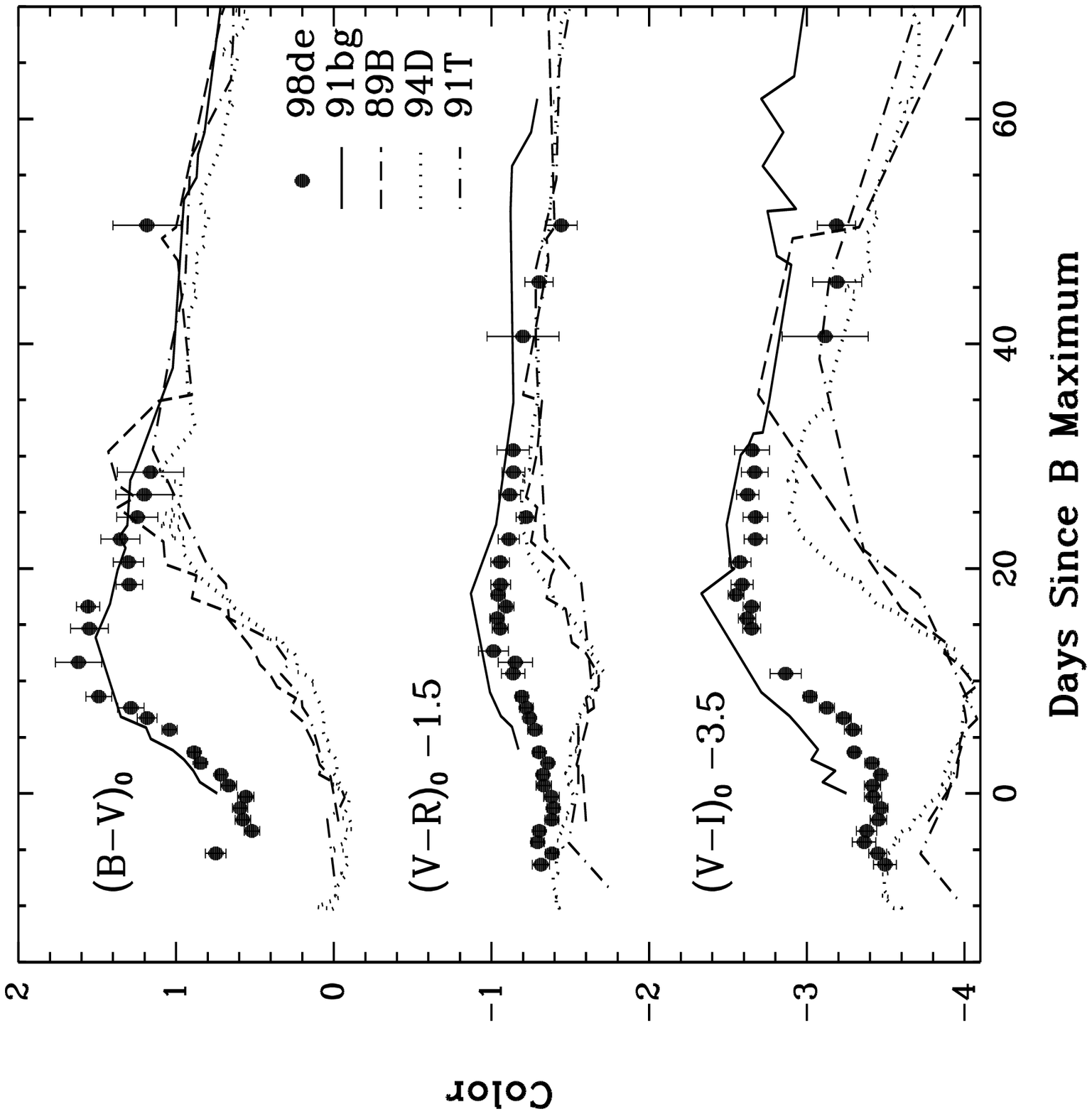}{3.4in}{-90}{50}{47}{-70}{290}}
{\footnotesize FIG. 5. --- The $(B - V)_0$, $(V - R)_0$, and $(V - I)_0$ color curves of SN~1998de, together with those of SN~1991bg (no reddening assumed), SN~1991T (dereddened by $E(B - V)$=0.13 mag), SN~1989B (dereddened by $E(B - V)$=0.37 mag), and SN~1994D (dereddened by $E(B - V)$=0.04 mag).  

}
\vspace{0.12in}

\begin{deluxetable}{cccccc}
\tablecaption{Absolute Peak Magnitudes of Several SNe Ia}
\tablehead{ 
\colhead{} & \colhead{$B$} & \colhead{$V$}  & \colhead{$R$} & \colhead{$I$}& \colhead{source, note}  
}
\startdata
    1994D   & $-$18.95 $\pm$ 0.18 & $-$18.90 $\pm$ 0.16 & $-$18.85 $\pm$ 0.15 & $-$18.68 $\pm$ 0.14 & 1,a \\
    1991bg  & $-$16.54 $\pm$ 0.32 & $-$17.28 $\pm$ 0.31 & $-$17.58 $\pm$ 0.31 & $-$17.68 $\pm$ 0.31 & 2,a \\
    1998de  & $-$16.95 $\pm$ 0.24 & $-$17.63 $\pm$ 0.21 & $-$17.79 $\pm$ 0.21 & $-$17.74 $\pm$ 0.12 & 3,b \\
\enddata
\tablenotetext{a}{Distance based on the surface brightness fluctuations method.}
\tablenotetext{b}{Magnitudes for $H_o=65$\kmsMpc.}
\tablerefs{(1) Richmond \etal~1995; (2) Turatto \etal~1996; (3) this paper.}
\end{deluxetable}

	We assume that there is negligible reddening due to the dust in the host galaxy of SN~1998de for the following reasons. SN~1998de is located 72\arcsec~from the nucleus of the NGC 252, in the outskirts of a lenticular (S0) galaxy, where there is generally very little dust. Moreover, the spectra exhibit no conspicuous \ion{Na}{1}~D line at the redshift of the host galaxy. Unfortunately, we do not have enough $B - V$ points after $t$~=~32 days in order to see whether the linear fit between intrinsic $B - V$ and time, which is described by Phillips \etal~(1999) (and whose sample also includes SN~1991bg), is also valid for our case. Nevertheless, we note that the late-time color evolution of $(V - R)_0$ and $(V - I)_0$, for which we have two more data points, seems to conform with the color evolution of normal SNe Ia, thus possibly indicating that there is very little host galaxy extinction.

	The prominent feature of the $(B - V)_0$ color evolution for both peculiar SNe (1991bg and 1998de) is their unusually red $(B - V)_0$ color at early epochs. At $B$ maximum SN~1998de has $(B - V)_0$=0.56~mag and SN~1991bg has $(B - V)_0$=0.69~mag, making SN~1998de about 0.6 mag redder than typical SNe Ia at this phase. Not only are the two peculiar SNe redder than normal SNe Ia, they also reach their reddest color (\ie, their maximum in a $B - V$ color curve) earlier than normal SNe Ia, at $t \approx$~13~days rather than at $t \approx$~30~days with $(B - V)_{0}^{max}$=1.45 mag for SN~1991bg and $(B - V)_{0}^{max}$=1.62~mag for SN~1998de. The $(V - R)_0$ color curve of SN~1998de shows a pattern of evolution which is again similar to that of SN~1991bg, but seems to be overall bluer by $\sim$0.12 mag. The region between $t = -5 $ and $t = 4$~days, for which there are no SN~1991bg data, seems to reveal a fairly stable color of $(V - R)_0\approx$~0.15 mag.

	The $(V - I)_0$ color evolution of SN~1998de seems to display an almost constant color of $(V - I)_0\approx$~0.05 mag between $t = -5 $ and $t = 4$~days. SN~1998de is 0.2$-$0.3~mag bluer than SN~1991bg at early times, but both objects reach their reddest color at about 17 days after $B$ maximum. Whereas the color curve of SN~1998de is distinct from those of normal and overluminous SNe during early times, it may approximate them for $t \geq$~40 days.

	Thus, the $(B - V)_0$, $(V - R)_0$, and $(V - I)_0$ color curves of SN~1998de generally have a similar pattern of evolution and are strikingly redder than those of normal and overluminous SNe Ia at early times ($t \la$~40 days). Each color curve has a maximum with the time differing only slightly: 13 days after $B$ maximum for the peak in $(B - V)_0$, and $t\approx17$~days for $(V - R)_0$ and $(V - I)_0$.

\section{ABSOLUTE MAGNITUDES}

	The light curves and color curves of SN~1998de suggest that it is subluminous, similar to SN~1991bg. In order to verify this, we need to find its absolute magnitude at maximum brightness and compare it with that of normal SNe Ia. Besides knowing the Galactic and extragalactic extinction, which was found in \S2.3, we need to accurately determine the distance to the host galaxy NGC~252 (UGC~491, PGC~2819). 

	NGC~252 is listed as the brightest member of group 12 of the Lyon Group of Galaxies (LGC) Catalog (Garcia 1993) and also as a member of the Perseus supercluster (Vettolani \& Baiesi Pillastrini 1987). Making use of the Lyon-Meudon Extragalactic Database\footnote{www-obs.univ-lyon1.fr} (LEDA), we obtain a mean heliocentric recession velocity of $cz$ = 4950\kms~for NGC~252. This estimate is close to those obtained by several other groups using different methods (\eg, neutral hydrogen line-width measurements by Eder, Giovanelli, \& Haynes 1991, optical measurements by Huchra, Vogeley, \& Geller 1999; see also de Vaucouleurs \etal~1991), all of which are between 4949 and 5023\kms.

	Since NGC~252 lies in the Hubble flow,\footnote{As indicated by Kraan-Korteweg (1986), we can neglect the influence of the Virgo cluster on the host galaxy of SN~1998de.} we can use the recession velocity and Hubble's law for deriving a reliable distance. Converting $v = cz$ to a velocity in the cosmic microwave background (CMB) frame according to the prescription of de Vaucouleurs \etal~(1991), we derive $v_{CMB} = 4625$\kms. Thus we find a distance of 71$\pm$6 ($H_o/65$) Mpc and a distance modulus of $\mu = (m-M) = 34.26 (\pm 0.20) - 5\textnormal{log}(H_o/65)$~mag, where the uncertainties reflect a possible peculiar motion of $\sigma_v = 420$\kms~of the host galaxy with respect to the Hubble flow (Davis 2000, private communication\footnote{The peculiar velocity was derived using predictions from the 1.2 Jy $IRAS$ survey and assuming $\beta=0.5$ in a linear perturbation theory.}).

	Using apparent peak magnitudes as listed in Table 3, together with our estimates of the extinction and distance, we derive the peak absolute magnitudes for SN~1998de in all filters (Table 4). The quoted uncertainties are the sums in quadrature of the uncertainties in peak magnitude, extinction, and distance. For comparison, we also list the absolute magnitudes of the subluminous SN~1991bg, as well as the normal SN Ia 1994D.

	Riess \etal~(1996, 1998a) describe an empirical method (the multicolor light curve shape; MLCS) to derive the absolute magnitude, distance, and total line-of-sight extinction of a Type Ia supernova. Applying the linear MLCS method to both light curves and color curves of SN~1998de, we derive a value of the ``luminosity correction'' $\Delta = 1.42\pm0.15$ mag and $A_V = 0$ mag. Given the uncertainties, this is indistinguishable from $\Delta = 1.44\pm0.10$ mag for SN~1991bg (Riess \etal~1996). The luminosity correction signifies the amount by which SN~1998de is dimmer in the $V$ band at the time of $B$-band maximum than a normal SN Ia with $<M_V>$ = $-$19.36~mag (on the absolute Cepheid distance scale). Thus, its absolute magnitude is $M_V$ = $- 17.96 \pm0.15$~mag at peak brightness in $V$. This value is 0.33 mag brighter than our more direct estimate, $M_V$ = $-17.63(\pm0.21) + 5\textnormal{log}(H_o/65)$~mag (Table 4), but considering the uncertainties in both methods, this is not a very significant discrepancy. MLCS also suggests that there is no extinction, while we assumed $E(B-V)$=0.06~mag of Galactic reddening. The MLCS method attempts to match the light curve of SN~1998de with that of the template, which is generated using a ``training set'' of SNe Ia. But since this training set consists primarily of normal SNe Ia and includes only one confirmed subluminous supernova (SN~1991bg) and a proposed one (SN~1986G; see, \eg, \flip), MLCS might be less reliable at low luminosities. Thus, we hesitate to adopt the luminosity of SN~1998de derived by Riess.\footnote{The ``two-parameter luminosity correction'' advocated by Tripp (1998) yields a corrected $M_B = -18.84$ mag for SN~1998de, significantly different from the value for normal SNe Ia ($M_B = -19.36$ mag, Riess \etal~1996). But this method is poorly calibrated for very red SNe Ia, and it does not distinguish between reddening by dust and intrinsically red colors. Thus, we do not adopt it here.}
	
	We can clearly discern from Table 4 that SN~1998de is a subluminous supernova and that it has magnitudes similar to those of SN~1991bg [on the surface-brightness-fluctuations (SBF) distance scale\footnote{Note that the SBF method yields a Hubble constant of $\sim$75\kmsMpc~(\eg, Tonry \etal~2000).}] in all four filters. These values are appreciably fainter and redder than the average, $M_B = -18.64(\pm0.05)+5\,\textnormal{log}(H_{o}/85) = -19.22$~mag and $M_V = -18.59(\pm0.06)+5\,\textnormal{log}(H_{o}/85) = -19.21$~mag for $H_o=65$\kmsMpc~as found by Vaughan \etal~(1995) for their \emph{ridge-line} SNe Ia (\ie, SNe Ia that suffered little extinction, but were also not exceptionally luminous). SN~1998de is even more extreme when compared to the Riess \etal~(1996) absolute magnitude template values ($M_V = -19.38$ and $M_B = -19.36$ mag) and the mean SN Ia magnitudes derived with the Cepheid distance calibration ($M_B = -19.65\pm0.13$ and $M_V = -19.60\pm0.11$ mag; see Saha \etal~1995). 

	The high quality of the data combined with our accurate distance measurement (excluding uncertainties in $H_{o}$) make SN~1998de an ideal candidate for testing current studies of the behavior of SNe Ia in the subluminous range. Riess et al. (1999b) determine the rise time (\ie, the time between explosion and maximum brightness) of nearby SNe Ia and find a strong correlation between the rise time, the postrise light-curve shape, and the peak luminosity. Applying the same method to SN~1998de, we find a rise time to $B$ maximum of $14.6 \pm 1.7$ days. This is very fast for a SN Ia (the fiducial SN Ia with \delm(B) = 1.1 mag and peak $M_V = -19.45$ mag has a rise time to $B$ maximum of $19.5 \pm 0.2$ days), but supports the trend that fast decliners are fast risers. Riess et al. emphasize that their conclusions are based on spectroscopically normal SNe Ia (Branch, Fisher, \& Nugent 1993), but they also seem valid for highly subluminous SNe Ia.

\vspace{0.4in}
{\plotfiddle{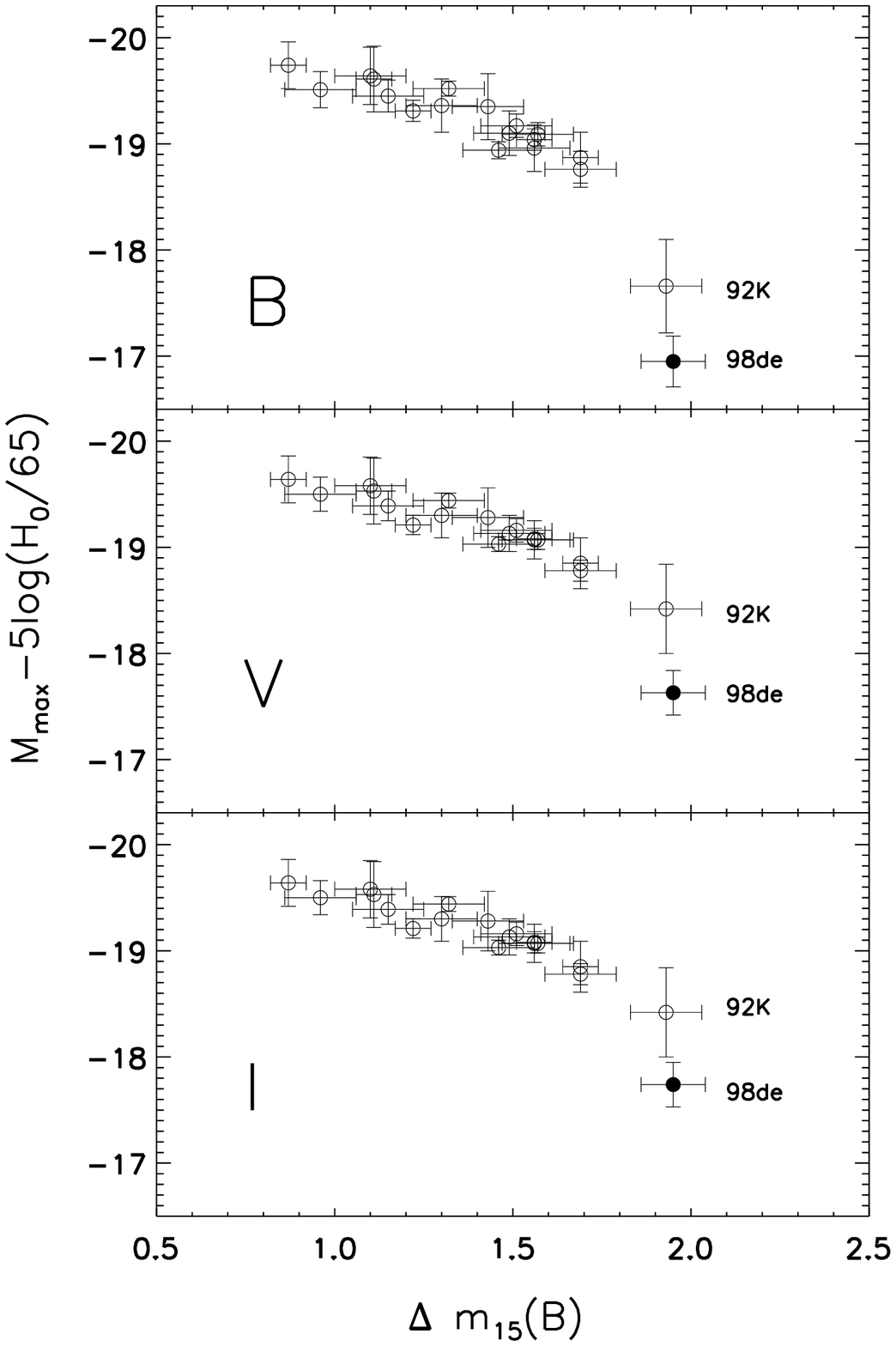}{4.9in}{0}{60}{60}{-50}{-40}}
{\footnotesize FIG. 6. --- Absolute $BVI$ magnitudes corrected for Galactic reddening plotted vs. the decline parameter \delm(B). The open circles indicate the 18 SNe Ia from Phillips \etal~(1999) with $z\geq0.01$ and non-significant host galaxy reddening; the filled circles indicate our data points for SN~1998de. 

}
\vspace{0.16in}


	One major attempt to accurately calibrate the peak luminosity of SNe Ia involves the peak luminosity versus decline rate relationship (Phillips 1993; Hamuy \etal~1996a,b;  Phillips \etal~1999). Using the sample consisting of 18 SNe Ia with $z\geq$0.01 and without significant host-galaxy reddening from Phillips \etal~(1999), we plot the absolute $BVI$ magnitudes corrected for Galactic reddening versus \delm(B) (Figure 6, open circles) and add our data points for SN~1998de (filled circles). Whereas SN~1992K seems to concur with the almost linear correlation (in fact, Phillips \etal~1999 find it to be quadratic), SN~1998de deviates from its expected behavior. However, more studies of subluminous SNe Ia are needed to accurately calibrate the luminosity versus decline rate relation at the subluminous end. 

\section{SPECTROSCOPY}
\subsection{Observations and Reductions}

	Spectroscopic observations were conducted with the Lick 3-m Shane telescope using the Kast CCD spectrograph (Miller \& Stone 1993) on the three dates listed in Table 5. The air mass was in all cases close to 1.0, and a slit width of 2\arcsec~was used.

\vspace{0.1in}
\begin{center}
\begin{tabular}{rrr}
\multicolumn{3}{c}{TABLE 5}\\
\multicolumn{3}{c}{Journal of Spectroscopic Observations\tablenotemark{a}}\\
\tableline
\tableline
UT Date & J.D.$-$  & Phase\tablenotemark{b} \\
DD/MM/YY &  $2,451,000.0$   & (days)\\
\tableline
31/08/98& 56.8  & $+31$ \\
20/09/98& 76.8  & $+51$\\
15/10/98& 101.7 & $+76$ \\
\tableline
\end{tabular}
\end{center}
{\footnotesize $^{\rm a}$
Lick 3-m reflector, 3,400$-$10,000 \AA.} \\
{\footnotesize $^{\rm b}$
Relative to the epoch of $B$ maximum (JD=2,451,026.3).} 
\vspace{0.1in}

	All optical spectra were reduced and calibrated employing standard techniques. To remove near-infrared fringing in the CCD, flatfield exposures were taken at the position of the supernova, as were He-Hg-Cd-Ne-Ar lamp exposures for wavelength calibration. The spectra were corrected for the continuum atmospheric extinction using mean extinction curves, and their telluric lines were removed following a procedure similar to that of Wade \& Horne (1988) and Bessell (1999). The final flux calibrations were derived from observations of one or more spectrophotometric standard stars (Oke \& Gunn 1983). All spectra shown in this paper have been corrected for reddening (as in \S2.3) and for their redshift.

	The spectral evolution of SN~1998de is illustrated in Figure 7. There was poor signal below 4000~\AA~and considerable fringing above 8000~\AA, affecting our observations more at late times, where the fringe amplitude was much closer to the supernova brightness. The sharp and narrow ``absorption lines'' in the $t =$ 76 d spectrum are noise spikes and cosmic rays, but we refrained from averaging or smoothing them in order to give an authentic impression of the quality of the data. 

	As mentioned in \S2.1, the spectrum reported by Garnavich \etal~(1998) and taken $\sim$6 days before $B$ maximum already shows the \ion{Ti}{2} absorption feature and a very strong \ion{Si}{2} feature at 5800~\AA, suggesting SN~1998de to be a SN~1991bg-type Ia event. Unfortunately, we have no spectra of our own around maximum light, with which we could have checked the spectral sequencing method described by Nugent et al. (1995), who find a correlation between the strength of certain spectral features of SNe Ia and their absolute $B$ magnitude. Our first spectrum was obtained 31 days after $B$ maximum and resembles that of the peculiar SN Ia 1991bg. It exhibits the broad trough at 4100$-$4500~\AA~identified as \ion{Ti}{2} (\flip; \tur; \maz) and shows signs of an early onset of the nebular phase. On the other hand, apparent evidence for abundance or excitation anomalies is present when compared with SN~1991bg.  

\vspace{0.6in}
{\plotfiddle{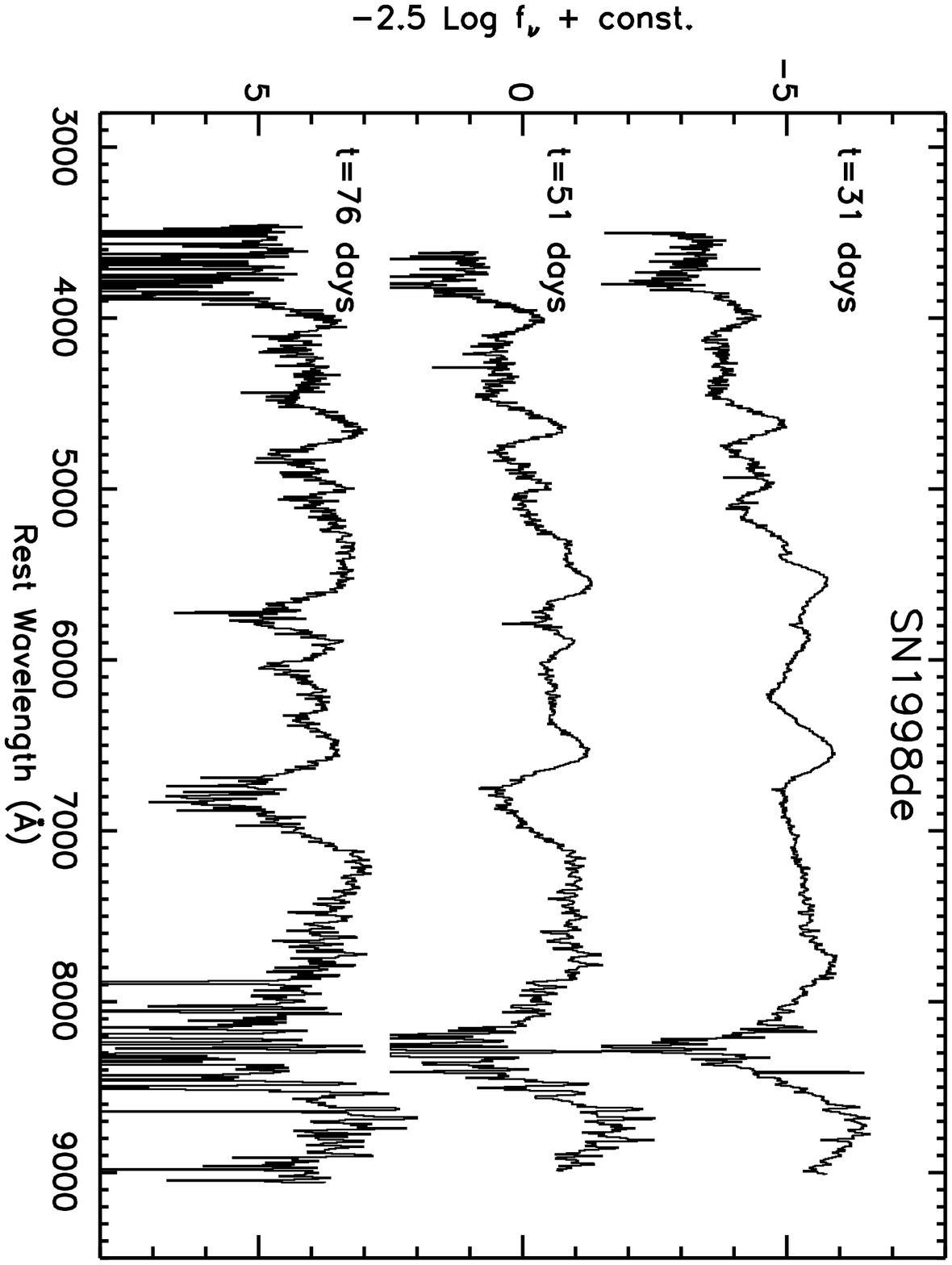}{2.8in}{90}{45}{47}{294}{-10}}
{\footnotesize FIG. 7. --- Montage of spectra of SN~1998de. The phases marked are relative to the date of $B$ maximum. To improve clarity, the spectra have been shifted vertically by arbitrary amounts. The effective units on the ordinate are magnitudes ($-$2.5 log$f_{\nu}$ + constant). Unless otherwise noted, all spectra in this paper are shown in the rest system of the host galaxy and corrections for reddening have been made. Note that the well-defined notch at the 5794 \AA~(5891 \AA~in the observed frame) is most probably due to improper subtraction of a night-sky emission line at an observed wavelength of 5893 \AA. 

}
\vspace{0.11in}

\subsection{The Spectrum at 31 Days after $B$ Maximum}

	In the $t$ = 31 d spectrum (Figure 8), Ca H\&K is seen in both absorption and emission as a classical P-Cygni profile, as expected for SNe Ia. However, whereas the emission component (at 3990~\AA) stays the same, the absorption part (at 3790 \AA) deepens as the object ages (Figure 7, although it is hard to discern its exact strength due to noise and the absence of any other emission features at shorter wavelengths), similar to the trend observed with SN~1991bg, where it grew even deeper. The broad absorption trough between 4120~\AA~and 4460~\AA, one of the most conspicuous and intriguing peculiarities of all very subluminous SNe Ia, such as SN~1991bg, has been identified with \ion{Fe}{3} by \lleib~by matching wavelengths, while \fflip, Ruiz-Lapuente \etal~(1993), and \mmaz~attribute it to a blend of \ion{Ti}{2} lines using spectral synthesis analysis. \fflip~and \mmaz~suggest that \ion{Mg}{2} \lam4481 might also contribute to the absorption. Since the evidence for \ion{Ti}{2} through the technique of spectral synthesis is more compelling, we adopt this particular interpretation. This trough contrasts sharply with the expected blend of Si, Mg, and Fe lines observed at comparable epochs in typical SNe Ia (\eg, Kirshner \etal~1993). Note that the trough in SN~1998de might be  somewhat broader (by $\sim$100~\AA) than the one in the SN~1991bg spectrum.

\vspace{0.4in}
{\plotfiddle{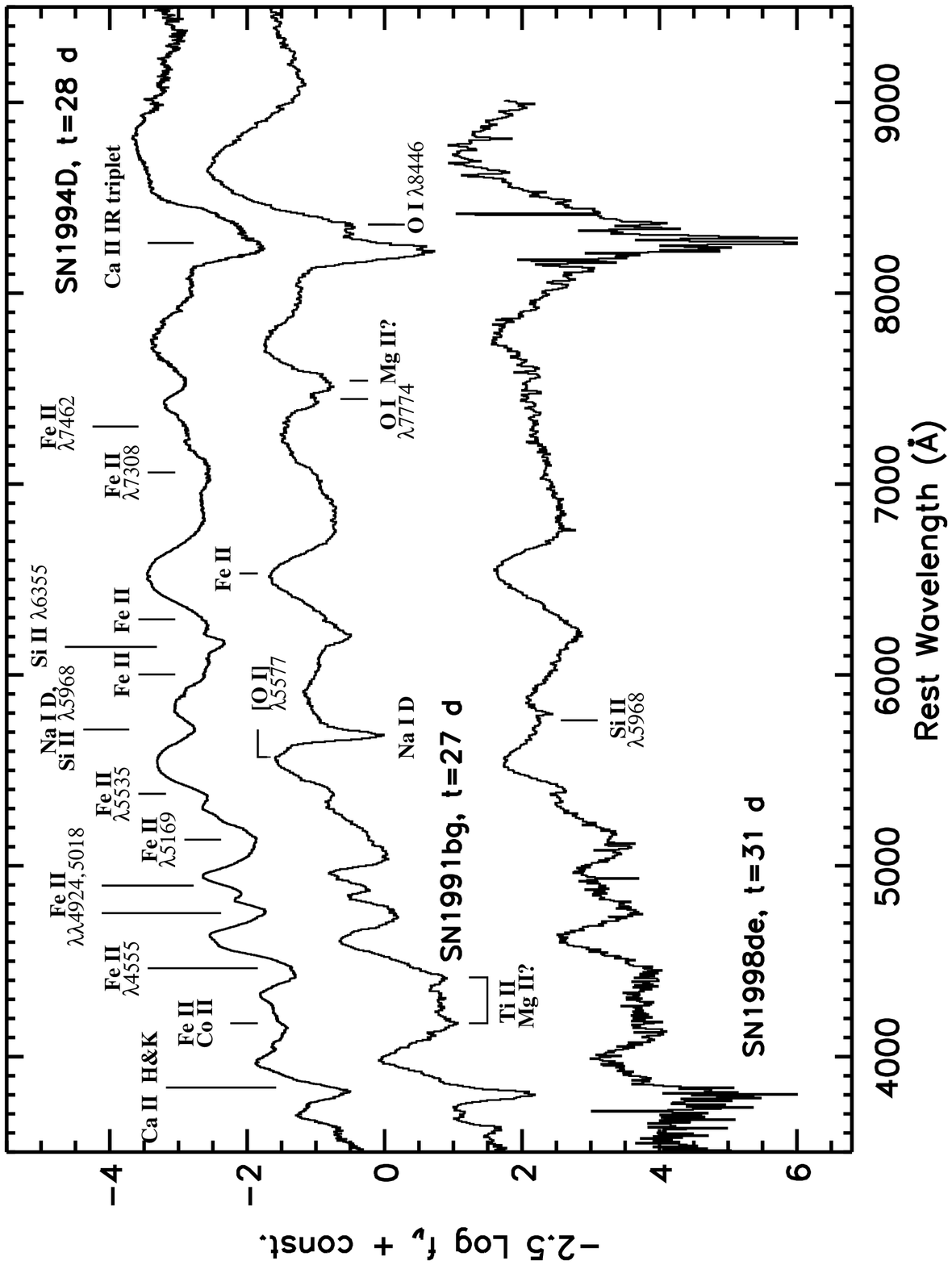}{2.8in}{-90}{45}{45}{-64}{260}}
{\footnotesize FIG. 8. --- The spectrum of SN~1998de about 4 weeks past maximum brightness, shown with comparable-phase spectra of SN~1991bg [no reddening assumed] and of the normal SN Ia 1994D [dereddened by $E(B - V)$=0.04 mag]. See text for sources of line identifications.

}
\vspace{0.15in}


	The wavelength range between 3800~\AA~and 5540~\AA~provides an impressive match to that of SN~1991bg. The emission line at about 5540~\AA~has been attributed to [\ion{O}{1}] \lam 5577 for SN~1991bg by \fflip~(see \leib~for an alternative identification), indicating the early start of the nebular phase. If we apply this identification also to SN~1998de, a significant fraction of the unburned oxygen is visible in dense states outside the bulk of the ejecta, as suggested by \mmaz~for SN~1991bg. \lleib~reject this interpretation, since the strength of the line and the absence of the [\ion{O}{1}] doublet at 6300~\AA~would imply very high densities ($n\approx10^{11}$ oxygen atoms $~{\rm cm^{\scriptscriptstyle -3}}$) and low temperatures ($T_e\leq2000$ K) for reasonable electron density ($n_{e}/n_{\textnormal{\tiny{O II}}} = 1$) and ionization ($n_{\textnormal{\tiny{O II}}}/n_{\textnormal{\tiny{O I}}} = 10^{-2}$; \leib). However, \mmaz~present compelling and intriguing evidence with the use of Monte Carlo simulations that SN~1991bg shows an indication of a high oxygen abundance due to incomplete burning and low temperatures. Their calculations may also explain the unusual strength of the \ion{Ti}{2} lines and the red color of SN~1991bg.

	Broad emission features, which are stronger in both SN~1991bg and SN~1998de than in SN~1994D and which are commonly observed in normal SNe Ia at later epochs ($t\ga$40 days), are the emission components of the Ca~H\&K feature and of the \ion{Ca}{2} near-infrared triplet at 8700~\AA, implying an early onset of the nebular phase.

	One of the most unusual characteristics displayed by the intermediate and late-time spectra of SN~1991bg is a very strong and relatively narrow absorption line measured near 5700~\AA, which increases in strength over time. As \ttur~show, this line, which is almost invisible in the maximum-light spectrum, persists for over 5 months with the line profile remaining nearly constant in time (FWHM~$\approx$~2500\kms). If we identify it with \ion{Na}{1}~D (\flip; see Fig.7 of \maz~for the reproduced \ion{Na}{1}~D line in a modeled spectrum), which indeed forms a broad absorption line in normal SNe Ia at this epoch, we obtain an expansion velocity of $v_{\textnormal{\tiny{Na I~D}}} $ = 9400\kms~for the absorbing layers. This is larger than the corresponding expansion velocity derived on the basis of the \ion{Si}{2} \lam6355 shift for the same epoch ($v_{\textnormal{\tiny{Si II}}}$ = 7000\kms). Both the narrow width and the high expansion velocity indicate that the element responsible for the 5700~\AA~feature is situated in a relatively thin outer layer (since the velocity of a shell is proportional to its radius at this phase). This could be evidence for abundance stratification in the ejecta of SN~1991bg as pointed out by \ttur. An alternative interpretation is given by \lleib, who propose that the 5700~\AA~feature is due to a blend of \ion{Si}{2} lines, but since it is very narrow, we hesitate to adopt this identification (although \ion{Si}{2} might be present, but contaminated by the very strong \ion{Na}{1}~D).

	Remarkably, the 5700~\AA~feature is not visible in any of the SN~1998de spectra, whereas other SN~1991bg-type objects like SN~1997cn (Turatto \etal~1998) and SN~1992K (Hamuy \etal~1994) display it. SN~1998de thus adds to the ``mystery'' of this conspicuous feature. It is interesting that the first major mismatch between SN~1998de and SN~1991bg occurs in the passband corresponding to the $V$ band. It is tempting to correlate the photometric mismatch found in the $V$ filter for $t$ = 31 days  (SN~1998de being $\sim$0.2 mag dimmer than SN~1991bg) with the 5700~\AA~feature and other, minor lines. 

	In SN~1998de, as in SN~1991bg, the primary characteristic feature of SNe Ia, \ion{Si}{2} \lam6355, seems to be contaminated by another absorption line. Unlike the case in SN~1991bg, however, both wings appear to be broadened owing to the presence of other elements. Instead of a narrow and well-defined core as found in normal SNe Ia such as SN~1994D, we see a V-shaped feature that fades as the supernova ages. Garnavich \etal~(1998) measure at $t = -6$ days the minimum of \ion{Si}{2} \lam6355~at 6185~\AA~and thus infer a photospheric expansion velocity of 13,300\kms~for SN~1998de, after correcting for the host galaxy's redshift. This is consistent with expansion velocities for normal SNe Ia which range between 11,000 and 13,000\kms~at maximum brightness (Branch, Drucker, \& Jeffery 1988; Barbon \etal~1990), and possibly with SN~1991bg, which showed a velocity of 9900\kms~at maximum light (\flip). Applying the same technique we find an expansion velocity of 7080\kms~for $t$ = 31 days, which is in excellent agreement with a velocity of $\sim$7000\kms~for SN~1991bg at $t = 27$ days and considerably smaller than the photospheric expansion velocity of SN 1994D ($v = 8970$\kms) at the same epoch.\footnote{Since the \ion{Si}{2} \lam 6355 line is not visible in later spectra, we cannot use it to obtain the expansion velocity, but by then the nebular phase has completely emerged and the photospheric model is no longer applicable.} These results are very interesting: whereas SN~1991bg showed a lower expansion velocity than normal SNe Ia, which led to the theoretical insights that the mass of the ejecta might be small for consistency with an early onset of the nebular phase and that complete burning was confined to the innermost regions (\tur; \maz), SN~1998de might not show such a clear trend. Again, early-time spectra would have been more useful for monitoring the development of the expansion velocity. 

	Note the broad emission near 6520~\AA, which usually begins to develop around $t$ = 2 weeks for normal SNe Ia and which is clearly visible at $t \approx 4$ weeks in the spectra of SNe 1994D, 1991bg, and SN~1998de. It remains present thereafter with some changes in shape and can be identified with \ion{Fe}{2} and later [\ion{Fe}{2}] (\leib; Filippenko 1997), and not with H$\alpha$ as one might erroneously infer. 

	A peculiarity which distinguishes SN~1998de from both SN~1991bg-like objects and normal SNe Ia is the nearly featureless region from 6800~\AA~to 7600~\AA~which lies in a wavelength range corresponding to the $I$ passband. The prominent absorption component of the \ion{O}{1} \lam 7774 P-Cygni feature and its contamination by \ion{Mg}{2} \lam 7877$-$7896 in the red wing as seen in SN~1991bg (\flip; \maz; \tur) and its twins (SNe 1997cn, 1992K) is not present, whereas the hump near 7760~\AA~is probably due to the emission component of \ion{O}{1} \lam7774 (but at a slightly longer wavelength when compared to SN~1991bg; \flip; \leib; \maz). The spectrum of SN~1998de may show a small indentation at the expected position of \ion{Mg}{2}, but the quality of the data is too poor for a definite conclusion. \mmaz~claim that the observed behavior of the \ion{O}{1} \lam 7774 absorption in SN~1991bg, which does not evolve redward (\ie, to lower velocities) during the first two weeks after $V$ maximum, is indicative of an outer oxygen shell containing unburned material from the progenitor, for which the observed [\ion{O}{1}] \lam5577 emission was the first clue. The absence of the \ion{O}{1} \lam 7774 absorption line in SN~1998de could be due to possible contamination by a blend of broad emission lines, but no likely candidates are known. A more feasible explanation could be that there is not as much [\ion{O}{1}] present in SN~1998de as in SN~1991bg. If so, perhaps SN~1998de does not have as much unburned oxygen left as did SN~1991bg.
 
	As in other SN Ia spectra, the obvious P-Cygni profile of the \ion{Ca}{2} near-infrared triplet is visible between 8000~\AA~and 8700~\AA, although both absorption and emission are much stronger than usual, similar to the appearance of SN~1991bg. Early-time spectra would be more conclusive for displaying the effects of color evolution --- indeed, we would expect those spectra to be redder than in normal SNe Ia. The consistently red color of SN~1991bg was interpreted by \mmaz~as an indication for a lower photospheric temperature, as would be expected with the presence of only incomplete nuclear burning. 

\subsection{The Spectra at 51 and 76 Days after $B$ Maximum}

	Figure 9 displays the spectra of SNe~1998de, 1991bg, and 1994D at about the same epoch of $t \approx$ 7 weeks, while Figure 10 shows the spectra of the same SNe at $t \approx$ 12 weeks. Our initial conclusion --- that differences between SN~1998de and SN~1991bg increase with increasing wavelength --- is confirmed. Besides the further emergence of emission lines, the following major changes have occurred in SN~1998de, and are clearly visible in Figure 10: 
(1) the [\ion{O}{1}] \lam 5577 and the \ion{O}{1} \lam7774 emission features have weakened and become broader;
(2) pronounced emission has appeared at 5900~\AA~whose position coincides with the [\ion{Co}{3}] emission in SN~1991bg (\leib; \tur; \maz); 
(3) a deep and broad absorption trough is present at $\sim$6800~\AA~adjacent to an emerging broad emission feature at $\sim$7200~\AA; and
(4) \ion{Si}{2} \lam 6355 has disappeared whereas \ion{Si}{2} \lam 5968 seems to have gained strength and may be contaminated in its blue wing.

	Most of the emission lines at this nebular stage are identified with forbidden lines of Fe and Co, the decay products of $^{56}$Ni, while the \ion{Ca}{2} emission is still strong in the spectra of both SN~1998de and SN~1991bg (Figure 10). 

\vspace{0.4in}
{\plotfiddle{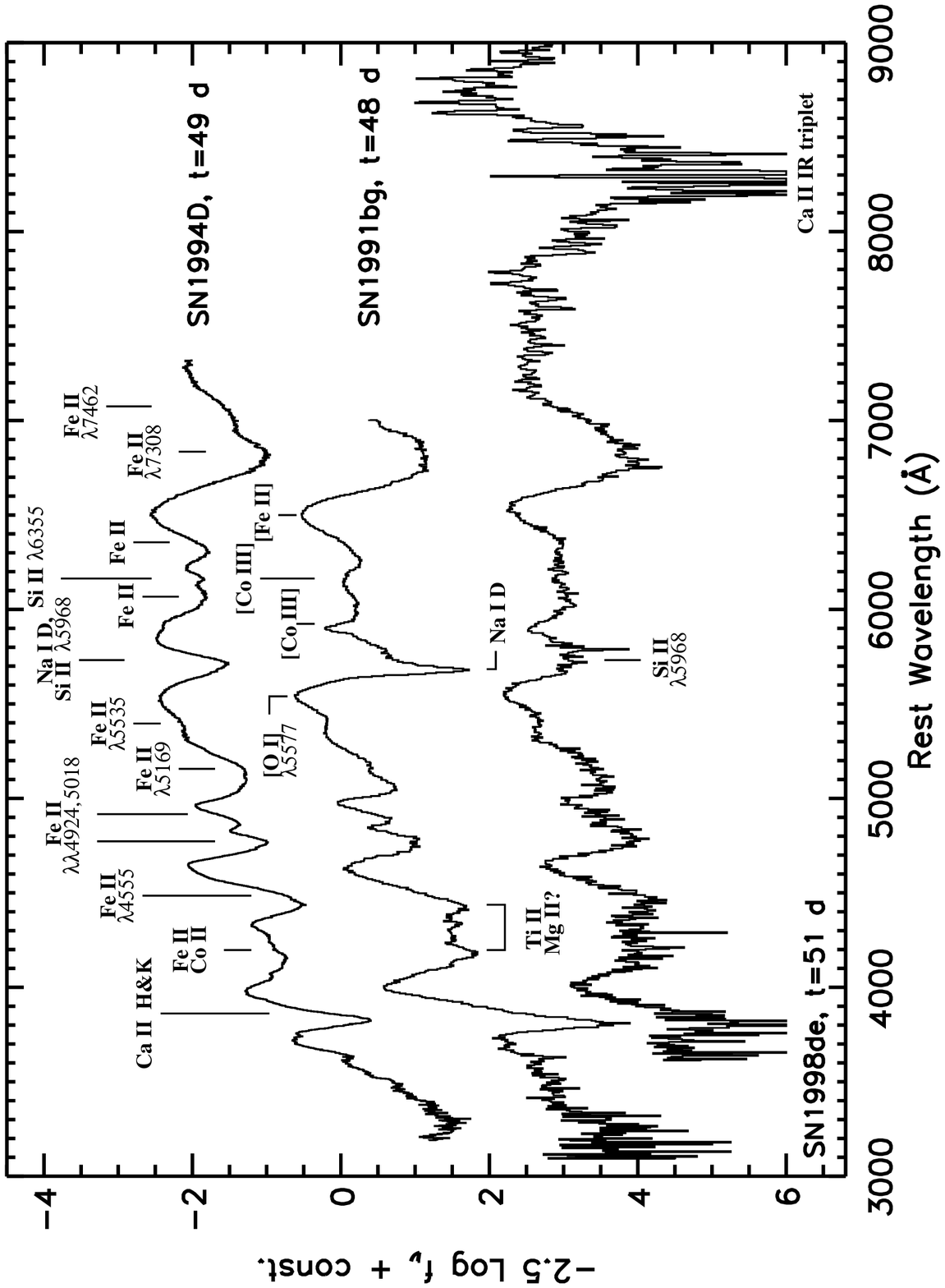}{2.8in}{-90}{45}{45}{-55}{260}}
{\footnotesize FIG. 9. ---  Same as in Fig. 8 for SNe Ia about 7 weeks past maximum.

}
\vspace{0.12in}

\vspace{0.4in}
{\plotfiddle{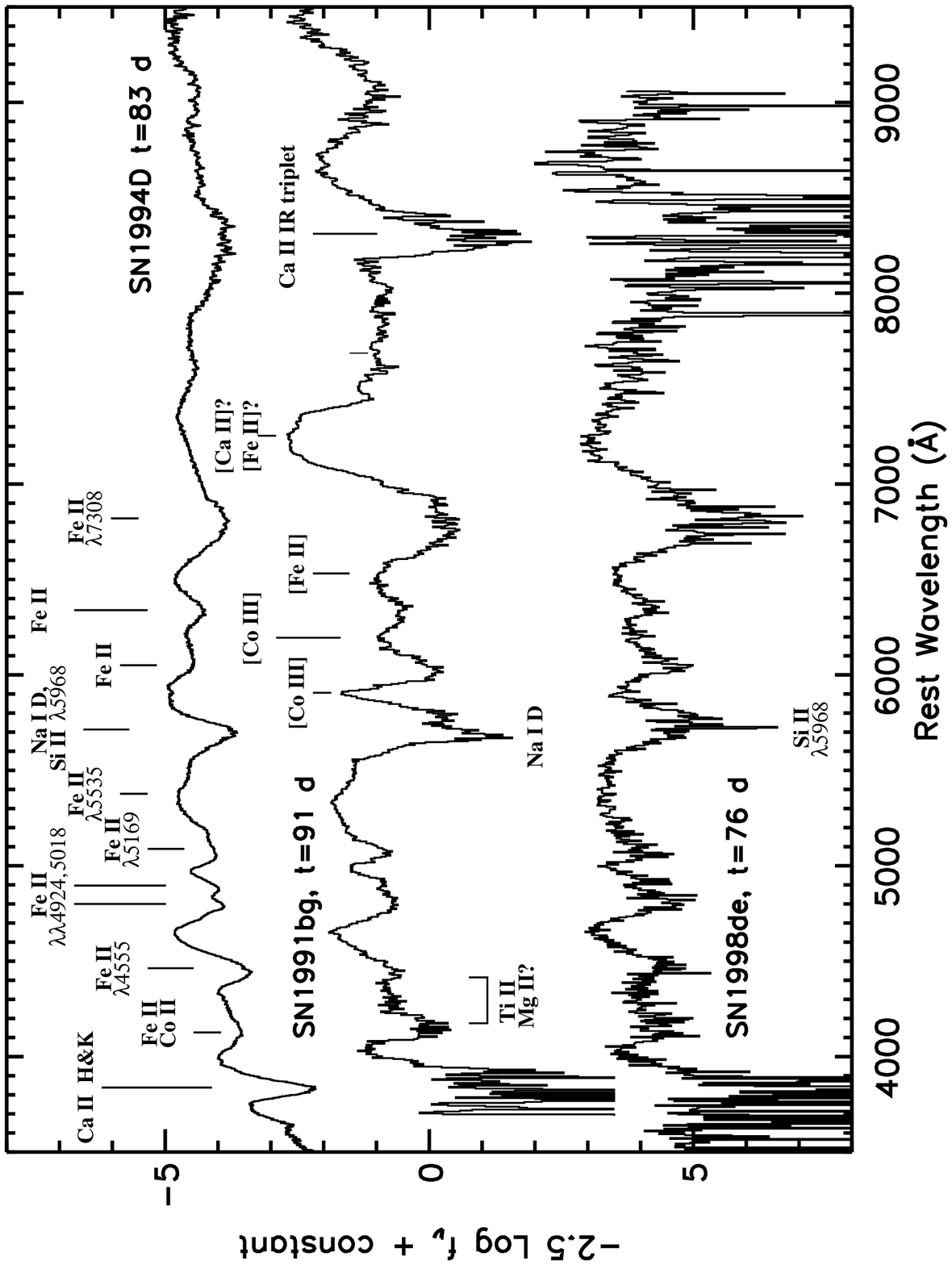}{2.8in}{-90}{45}{45}{-55}{260}}
{\footnotesize FIG. 10. ---  Same as in Fig. 9 for SNe Ia about 12 weeks past maximum.

}
\vspace{0.12in}

	The oxygen line deterioration can be explained by the fact that the photosphere is receding into the central parts of the ejecta, where the higher-mass elements have been produced. The drifting of emission lines to the red, most notably the [\ion{Fe}{3}] feature at 4615 \AA, illustrates the decrease in the expansion velocity as we look deeper into the ejecta. 
The broad absorption at 6800~\AA~is most probably due to \ion{Fe}{2}, whose signature is visible in SN~1991bg as well as in SN~1994D (and generally in SNe Ia). The adjacent emission at 7200~\AA, which is only visible at $t = 76$ days, could be tentatively identified as a blend of [\ion{Ca}{2}] \lam \lam 7291, 7324 (\flip; Ruiz-Lapuente \etal~1993) or a blend of [\ion{Fe}{2}] (\leib). Note that the corresponding feature in SN~1991bg is observed at a slightly longer wavelength (7250~\AA) and that it dominates the spectrum of SN~1991bg, whereas it is very broad and not as distinct in the spectrum of SN~1998de (also the red shoulder seems to be contaminated), and is absent in the case of SN~1994D.

	Since the \ion{Si}{2} \lam 5968 absorption is not contaminated by a \ion{Na}{1}~D line, it is easily visible as a distinct feature at nearly the same position as in the $t$ = 31 d spectrum, only deeper. Because this \ion{Si}{2} feature deepens between the two last epochs and is accompanied by a receding photosphere which reveals the deeper layers, we could have evidence that nuclear burning was incomplete in these layers.

	The dominant and narrow [\ion{Co}{3}] emission feature of SN~1991bg visible at 5908 \AA~has been regarded by \mmaz~as an indication of a centrally peaked distribution of $^{56}$Ni and as the result of complete incineration of matter in the central part of the ejecta, assuming that the emissivity is proportional to the Co abundance and to the local gamma-ray deposition rate in the Co-rich material. Combining this with their observation of oxygen during the photospheric phase, \mmaz~suggest that complete burning did not take place everywhere, and that the outer layers underwent incomplete burning in contrast to the inner part. For SN~1998de, the corresponding [\ion{Co}{3}] emission is observed at nearly the same position of around 5900~\AA~in both the $t=$ 51 d and the $t=$ 76 d spectra; however, it is not as narrow as in SN~1991bg. This might indicate that complete incineration extended to a larger radius in SN~1998de than it did in the case of SN~1991bg. [\ion{Co}{3}] \lam \lam 6129, 6197 emission is also found at around 6210~\AA~in the $t=$ 76 d spectrum, matching the features at 6144~\AA~and 6204~\AA~in SN~1991bg (\tur). Furthermore, we see the afore-mentioned emission of [\ion{Fe}{2}] (\leib; see \tur~for alternative identification) at around 6520~\AA~in the spectrum of SN~1998de, a feature which is found in all three SNe (1994D, 1991bg, 1998de) and which was already visible in their intermediate-phase spectra (see Figure 8). 

	The \ion{Ti}{2} trough between 4100~\AA~and 4500~\AA, a primary characteristic of SN1991bg-type objects, is still distinctly visible at these late times. Thus, this feature can be used as a powerful tool for distinguishing between peculiar and normal SNe Ia at all epochs, even for SNe at large redshifts (whose restframe visible and red spectral regions are shifted into the range where the infrared fringing of CCDs affects the data). The strong presence of \ion{Ti}{2} can be explained either in terms of abundance anomalies or by excitation differences. Both \fflip~and \mmaz~favor the interpretation that the red color and hence low excitation temperature is largely responsible for producing the observed strong \ion{Ti}{2} lines in local thermodynamic equilibrium (see, \eg, Fig. 1 of Branch 1987) and, simultaneously, the higher-excitation lines of the other ions like \ion{Ca}{2}.

\section{DISCUSSION}
	In SNe Ia, the energy of the initial explosion gained through thermonuclear fusion is used up by the adiabatic expansion of the ejecta (\eg, Leibundgut \&~Pinto 1992), whereas the light curve is powered by the decay of radioactive $^{56}$Ni and $^{56}$Co. It is also believed that there is a direct link between the mass of the \nick~produced during the explosion and the peak bolometric luminosity (see, \eg, Arnett, Branch, \&~Wheeler 1985; H\"oflich \etal~1996). The observed low $optical$ luminosity of SN~1998de is indicative of its bolometric luminosity for the following reasons: the peak of the energy distribution of SN~1998de lies in the optical part of the spectrum, thus we can be confident that the low observed optical luminosity translates into a similarly low bolometric luminosity. Furthermore, there is no plausible way to have a large UV excess, nor is there evidence for a large amount of dust, which could have reprocessed the optical photons of SN~1998de to provide a significant infrared excess.

	Therefore, we can conclude that the mass of the synthesized nickel was smaller in SN~1998de than in normal SNe Ia. \mmaz~derive a value of $\sim$0.07 \sm~of synthesized $^{56}$Ni for SN~1991bg, which agrees well with the original estimate of Ruiz-Lapuente \etal~(1993) of 0.05 to 0.1 \sm~and which is a factor of 4 to 8 smaller than in normal SNe Ia. We could expect that similar amounts of nickel have been produced in both SN~1991bg and SN~1998de, as both SNe exhibit similar absolute magnitudes in all filters. However, spectral synthesis calculations and bolometric light curves are necessary for obtaining a definitive value for the nickel mass. 

	On the other hand, there are a number of indications that suggest that SN~1998de might have experienced a slightly more powerful explosion than SN~1991bg did. Among these are the marginally slower decline of the light curves in the $V$, $R$, and $I$ bands and the offset of the color curves toward the blue. The fact that SN~1998de appears to show the same luminosities when compared with SN~1991bg (on the same distance scale) is only mildly inconsistent with the other pieces of evidence. 

	The models which have been proposed to explain the physics of SNe Ia in the last 30 years can be divided into two main categories: those that describe different explosion mechanisms with the same progenitor system, and those that propose different systems as the reason for the heterogeneity of SNe Ia and in particular as production sites of subluminous SNe Ia. Despite the partial success of various models (see reviews by, \eg, Woosley \&~Weaver 1986; Wheeler \& Harkness 1990; Branch \etal~1995; Livio 2000; Hillebrandt \& Niemeyer 2000), no general consensus about the exact character of the explosion mechanism and the evolutionary track that leads to the SN Ia event has yet been reached.

	H\"{o}flich, Khokhlov, \& Wheeler (1995) used the pulsating class of delayed detonation models to establish the principle that Chandrasekhar mass C-O white dwarfs give a common explosion mechanism accounting for both normal $and$ subluminous SNe Ia, with SN~1991bg as their example for the latter group. However, as \ttur~correctly point out, their model requires a large reddening for SN~1991bg [$E(B - V) = 0.30$ mag] in order to harmonize the theoretical luminosity with the observed value, while observations indicate a much smaller value close to zero (\flip; \leib; \tur). Also, the decline rates of their SN~1991bg models are similar to those of their normal SN Ia models, while SN~1991bg is observed to be much faster.

	As mentioned by \fflip~and others, an alternative scenario for subluminous SNe Ia might involve ejecta with mass appreciably $smaller$ than the Chandrasekhar limit that would account for several of the peculiar features of SN~1998de. The trapping of gamma rays produced by the decay of $^{56}$Ni and $^{56}$Co would be less effective due to the low column depth of the ejecta, thus leading to the rapid decline in the observed $BVRI$ light curves after maximum light and possibly to the low optical luminosity. Also, low-mass ejecta expanding at only a slightly slower velocity than in normal SNe Ia would allow the photosphere to recede more quickly than usual, revealing the deeper layers and giving rise to the early emergence of the nebular spectrum. 

	The decline in photospheric expansion velocity from $\sim$13,000\kms~at $t = -6$ days (Garnavich \etal~1998) to 7080\kms~at $t = $ 31 days might be consistent with that of SN~1991bg, which showed an expansion velocity of $\sim$7000\kms~at $t =$ 27 days. Unfortunately, there are no other data points available for SN~1998de, especially since photospheric velocities are not defined during the nebular phase (\ie, for our last two spectra), and no premaximum spectra are available for SN~1991bg. 

	The consistently small velocity observed in SN~1991bg was interpreted by \ttur~as a possible indication that complete nuclear burning was confined to the innermost regions. Other phenomena which are thought to support this conclusion are the high abundance of oxygen (suggested by the [\ion{O}{1}] \lam5577 and \ion{O}{1} \lam7774 lines) left over from the progenitor (\tur; \maz), and the narrow [\ion{Co}{3}] emission at 5908 \AA. Using the presence and strength of those features as observed in SN~1998de, we might infer that the inner region of complete burning was somewhat larger in SN~1998de than in SN~1991bg; indicators include the weaker appearance of \ion{O}{1} \lam7774 and the somewhat broader [\ion{Co}{3}] emission at 5900~\AA~in SN~1998de. Furthermore, the hypothesis that SN~1998de might have experienced a slightly more powerful explosion than SN~1991bg did is consistent with this model. The red color of both SNe 1998de and 1991bg seems to implicate a low excitation temperature which gives rise to the \ion{Ti}{2} trough found at the blue end of the spectrum without requiring an overabundance of titanium. 

	Here, we briefly review some of the scenarios proposed to explain the ejection of a sub-Chandrasekhar envelope. Livne (1990), Woosley \& Weaver (1994), and H\"{o}flich \& Khokhlov (1996) suggest that the $progenitor$ itself has a mass below the Chandrasekhar limit. According to their models, a sub-Chandrasekhar-mass C-O white dwarf detonates due to the ignition of an outer He shell accumulated on top of it. Their predictions include a lower luminosity, a lower production rate of nickel, and an overabundance of titanium. In addition, the observed light curves can be adequately reproduced with this model. The absence of unburned He in the spectra might be explained by the low excitation temperatures present. On the other hand, they predict bluer colors at maximum brightness, while SNe 1991bg and 1998de are very red at that phase, as well as an outer Ni shell, which has not been observed in any of the subluminous SNe Ia. Moreover, they do not account for the outer oxygen shell.

	Another possibility might be a double-degenerate system of intermediate-mass C-O white dwarfs that merge, giving rise to the disruption of the less massive star, followed by the formation of a thick, hot accretion disk around the more massive companion and the subsequent explosion of the remaining white dwarf when it reaches the Chandrasekhar limit (\eg, Iben \& Tutukov 1984; Webbink 1984). As a consequence only a small mass of $^{56}$Ni would be produced, and there would be a mixture of burned and unburned material as observed in SN~1991bg and SN~1998de. However, the outcome of such a coalescence is still not completely understood; there are strong indications that it may rather lead to accretion-induced collapse forming a neutron star than to a SN event (see, \eg, Saio \& Nomoto 1985, 1998; Woosley \& Weaver 1986; Mochkovitch \& Livio 1990; Mochkovitch, Guerrero, \& Segretain 1997).

	Unfortunately, none of the proposed models unambiguously emerges as the preferred candidate for explaining subluminous SNe Ia, since each of the models appears beset by some problems as discussed above. Detailed studies of more subluminous SNe are needed to provide additional constraints for the models. Indeed, the ongoing efforts of LOSS to conduct photometric follow-up observations of recently discovered subluminous SNe Ia promise to contribute significantly; the results will be discussed by Li \etal~(2001c).
 
\section{CONCLUSIONS}
	
	We have presented a set of photometric and spectroscopic observations of the recent SN~1998de, which reveal it to be an intrinsically faint SN Ia, similar to, but not identical with SN~1991bg. Similarities include
(1) the intrinsically low luminosity;
(2) the remarkably red colors near maximum;
(3) the narrow peak in all bands;
(4) the rapid and monotonic decline of the light curves, including in the $I$ band;
(5) the strong \ion{Ti}{2} absorption lines throughout all observed epochs;
(6) the early transition to a nebular spectrum; and 
(7) the strong [\ion{Ca}{2}] emission lines at late times. The very good coverage and high quality of the $BVRI$ light curves make SN~1998de a defining template for subluminous SNe Ia, superior to the conventional template SN~1991bg, and clearly establish SN~1998de as a fast riser. 

	Whereas the $B$ passband provides the best early-time match between SN~1991bg and SN~1998de, the deviations in both light curves and spectra between the two SNe Ia seem to increase toward longer (redder) wavelengths. Features of SN~1998de that distinguish it from SN~1991bg include
(1) a plateau phase (or a possible second maximum) in the $I$ band, very shortly after the first maximum;
(2) a slower decline of the light curves in the $V$, $R$, and $I$ bands;
(3) slightly bluer colors at intermediate phases; 
(4) the absence of the conspicuous \ion{Na}{1}~D absorption found at 5700~\AA~in the spectra of SN~1991bg; and
(5) the evolution of a region (6800$-$7600~\AA) in the spectra of SN~1998de from featureless to feature-rich.
Our comparison of SN~1998de and SN~1991bg may imply that subluminous SNe Ia are not fully described as a one-parameter [\delm($B$)] family. The somewhat slower decline rates and bluer colors, combined with the evidence that the region of complete burning was larger in SN~1998de than in SN~1991bg, suggest that SN~1998de might have experienced a slightly more powerful explosion than SN~1991bg did. 
 	The observed peculiarities of SN~1998de (and of SN~1991bg) might be explained in terms of ejecta with a mass smaller than that found in normal SNe Ia. However, at this time Chandrasekhar-mass models seem to provide an equally acceptable explanation.

	Our analysis strengthens the conclusion that SN~1991bg was not a unique event, as already proposed by studies of several other intrinsically dim SNe Ia. Further conclusions to be drawn --- besides the insights into the intrinsic workings of SNe Ia --- concern the implications of our observations for the calibrated-candle assumption of SNe Ia and for their usage as cosmological distance indicators. We propose that intrinsically subluminous SNe Ia should be identified in samples of SNe Ia used for cosmological purposes, with the \ion{Ti}{2} trough at the blue end of the spectrum and the absence of a clear secondary maximum in the restframe $I$ band as possible criteria. It is advisable that the standard calibration methods be applied with caution on such objects; for example, the ``snapshot'' method of deriving distances of SNe Ia (Riess \etal~1998b) currently assumes that the objects are normal.

\acknowledgments

	The work of A. V. F.'s group at U. C. Berkeley is supported by National Science Foundation grants AST-9417213 and AST-9987438. KAIT was made possible by generous donations from Sun Microsystems, Inc. (Academic Equipment Grant Program), Photometrics Ltd., the Hewlett-Packard Company, AutoScope Corporation, Lick Observatory, the National Science Foundation, the University of California, and the Sylvia and Jim Katzman Foundation. We thank the referee, Brian P. Schmidt, for many useful suggestions.


\begin{thebibliography}{}

\bibitem[Arnett \etal~(1985)]{arne85}Arnett, W. D., Branch, D., \& Wheeler, J. C. 1985, \nat, 314, 337
\bibitem[Barbon \etal~(1990)]{barb90}Barbon, R., Benetti, S., Cappellaro, E., Rosino, L., \& Turatto, M. 1990, \aap, 237, 79
\bibitem[Bessell (1999)]{bess99}Bessell, M. S. 1999, \pasp, 765, 1426
\bibitem[Branch (1987)]{bran87}Branch, D. 1987, \apjl, 320, L121
\bibitem[Branch (1998)]{bran98}---------. 1998, \araa, 36, 17
\bibitem[Branch \etal~(1988)]{bran88}Branch, D., Drucker, W., \& Jeffery, D. J. 1988, \apjl, 330, L117
\bibitem[Branch \etal~(1993)]{bran93}Branch, D., Fisher, A., \& Nugent, P. 1993, \aj, 106, 2383
\bibitem[Branch \etal~(1988)]{bran95}Branch, D., Livio, M., Yungelson, L. R., Boffi, F. R., \& Baron, E. 1995, \pasp, 197, 1019
\bibitem[Branch \etal~(1992)]{bran92}Branch, D., \& Tammann, G. A. 1992, \araa, 30, 359 
[Cousins (1981)]{cous81}Cousins, A. W. J. 1981, South African Astron. Obs. Circ., 6, 4
\bibitem[de Vaucouleurs\etal~(1991)]{vauc91}de Vaucouleurs, G., \etal~1991, Third Reference Catalogue of Bright Galaxies (New York: Springer)
\bibitem[Eder \etal~(1991)]{eder91}Eder, J., Giovanelli, R., \& Haynes, M. P. 1991, \aj, 102, 572
\bibitem[Filippenko (1997)]{flip97}Filippenko, A. V. 1997, \araa, 35, 309
\bibitem[Filippenko \etal~(1986)]{flip86}Filippenko, A. V., Porter, A. C., Sargent, W. L. W., \& Schneider, D. P. 1986, \aj, 92, 1341
\bibitem[Filippenko \etal~(1992a)]{flip92a}Filippenko, A. V., \etal~1992a, \apj, 384, L15
\bibitem[Filippenko \etal~(1992b)]{flip92b}---------. 1992b, \aj, 104, 1543 (F92)
\bibitem[Filippenko \etal~(2001)]{flip01}---------. 2001, PASP, in preparation
\bibitem[Ford \etal~(1993)]{ford93}Ford, C. H., \etal~1993, \aj, 106, 1101
\bibitem[Garcia (1993)]{garc93}Garcia, A. M. 1993, A\&AS, 100, 47
\bibitem[Garnavich \etal~(1998)]{garn98}Garnavich, P., Jha, S., \& Kirshner, R. 1998, \iaucirc~6980
\bibitem[Hamuy \etal~(1996a)]{hamu96a}Hamuy, M., Phillips, M. M., Suntzeff, N. B., Schommer, R. A., Maza, J., \& Avil\'{e}s, R. 1996a, \aj, 112, 2391
\bibitem[Hamuy \etal~(1996b)]{hamu96b}Hamuy, M., Phillips, M. M., Suntzeff, N. B., Schommer, R. A., Maza, J., Smith, R. C., Lira, P., \& Avil\'{e}s, R. 1996b, \aj, 112, 2438
\bibitem[Hamuy \etal~(1993)]{hamu93}Hamuy, M., Phillips, M. M., Wells, L. A., \& Maza, J. 1993, \pasp, 105, 787
\bibitem[Hamuy \etal~(1994)]{hamu94}Hamuy, M., \etal~1994, \aj, 108, 2226
\bibitem[Harkness \& Wheeler~(1990)]{hark90}Harkness, R. P., \& Wheeler, J. C. 1990, in Supernovae, ed. A. G. Petschek (New York: Springer Verlag), p. 1
\bibitem[Hillebrandt \& Niemeyer (2000)]{hill00}Hillebrandt, W., \& Niemeyer, J. C. 2000, \araa, 38, 191
\bibitem[H\"{o}flich, \& Khokhlov (1996)]{hoef96}H\"{o}flich, P., \& Khokhlov, A. M. 1996, \apj, 457, 500
\bibitem[H\"{o}flich, Khokhlov, \& Wheeler (1995)]{hoef95}H\"{o}flich, P., Khokhlov, A. M., \& Wheeler, J. C. 1995, \apj, 444, 831
\bibitem[H\"{o}flich \etal~(1996)]{hoefetal96}H\"{o}flich, P., Khokhlov, A. M., \& Wheeler, J. C., Phillips, M. M., Suntzeff, N. B., \& Hamuy, M. 1996, \apj, 472, L81
\bibitem[Huchra \etal~(1999)]{huch99}Huchra, J. P., Vogeley, M. S., \& Geller, M. J. 1999, \apjs, 121, 287
\bibitem[Iben \& Tutukov (1984)]{iben84}Iben, I., Jr., \& Tutukov, A. V. 1984, \apjs, 54, 335
\bibitem[Jeffery \etal~(1992)]{jeff92}Jeffery, D. J., Leibundgut, B., Kirshner, R. P., Benetti, S., Branch, D., \& Sonneborn, G. 1992, \apj, 397, 304
\bibitem[Jha \etal~(1999)]{jha99}Jha, S., \etal~1999, \apjs, 125, 73 
\bibitem[Johnson \etal~(1966)]{john66}Johnson, H. L., Mitchell, R. I., Iriarte, B., \& Wisniewski, W. Z. 1966, Commun. Lunar Planet. Lab., 4, 99
\bibitem[Kirshner, \etal~(1993)]{kirs93}Kirshner, R. P., \etal~1993, \apj, 415, 589
\bibitem[Kraan-Korteweg (1986)]{kraa86}Kraan-Korteweg, R. C. 1986, \aapr, 66, 255
\bibitem[Landolt (1992)]{land92}Landolt, A. U. 1992, \aj, 104, 340
\bibitem[Leibundgut \& Pinto (1992)]{leib92}Leibundgut, B., \& Pinto, P. A. 1992, \apj, 401, L49 
\bibitem[Leibundgut \etal~(1993)]{leib93}Leibundgut, B., \etal~1993, \aj, 105, 301 (L93)
\bibitem[Li \etal~(2001a)]{li01a}Li, W. D., Filippenko, A. V., \& Riess, A. G. 2001a, \apj, in press  
\bibitem[Li \etal~(2001b)]{li01b}Li, W. D., Filippenko, A. V., Treffers, R. R., Riess, A. G., Hu, J., \& Qiu, J. 2001b, \apj, in press 
\bibitem[Li \etal~(2000a)]{li00a}Li, W. D., \etal~2000a, in Cosmic Explosions, eds. S. S. Holt and W. W. Zhang (New York: American Institute of Physics), p. 103
\bibitem[Li \etal~(2000b)]{li00b}Li, W. D., \etal~2000b, in Cosmic Explosions, eds. S. S. Holt and W. W. Zhang (New York: American Institute of Physics), p. 91
\bibitem[Li \etal~(2001c)]{li01c}Li, W. D., \etal~2001c, in preparation
\bibitem[Livio (2000)]{livi00}Livio, M. 2000, in Type Ia Supernovae: Theory and Cosmology, ed. J. C. Niemeyer \& J. W. Truran (Cambridge: Cambridge University Press), p.33
\bibitem[Livne (1990)]{livn90}Livne, E. 1990, \apj, 354, L53  
\bibitem[Mazzali \etal~(1997)]{mazz97}Mazzali, P. A., Chugai, N., Turatto, M., Lucy, L. B., Danziger, I. J., Cappellaro, E., Della Valle, M., \& Benetti, S. 1997, MNRAS, 284, 151 (M97)
\bibitem[Miller \& Stone (1993)]{mill93}Miller, J. S., \& Stone, R. P. S. 1993, Lick Obs. Tech. Rep. 66 (Santa Cruz: Lick Obs.)
\bibitem[Mochkovitch \etal~(1997)]{moch97}Mochkovitch, R., Guerrero, J., \& Segretain, L. 1997, in Thermonuclear Supernovae, eds. P. Ruiz-Lapuente, R. Canal, \& J. Isern (Dordrecht: Kluwer), p. 187
\bibitem[Mochkovitch \& Livio (1990)]{moch90}Mochkovitch, R., \& Livio, M. 1990, \aap, 236, 378
\bibitem[Modjaz \etal~(1998)]{modj98}Modjaz, M., \etal~1998, \iaucirc~6977
\bibitem[Nugent \etal~(1995)]{nuge95}Nugent, P., Phillips, M., Baron, E., Branch, D., \& Hauschildt, P. 1995, \apj, 455, L147
\bibitem[Oke \& Gunn (1983)]{oke83}Oke, J. B., \& Gunn, J. E. 1983, \apj, 266, 713
\bibitem[Pagel (1997)]{page97}Pagel, B. E. J. 1997, Nucleosynthesis and Chemical Evolution of Galaxies (Cambridge: Cambridge University Press)
\bibitem[Perlmutter \etal~(1997)]{perl97}Perlmutter, S., \etal~1997, \apj, 483, 565
\bibitem[Perlmutter \etal~(1999)]{perl99}---------. 1999, \apj, 517, 565
\bibitem[Phillips (1993)]{phil93}Phillips, M. M. 1993, \apj, 413, L105
\bibitem[Phillips \etal~(1992)]{phil92}Phillips, M. M., Wells, L. A., Suntzeff, N. B., Hamuy, M., Leibundgut, B., Kirshner, R. P., \& Foltz, C. B. 1992, \aj, 103, 1632
\bibitem[Phillips \etal~(1999)]{phil99}Phillips, M. M., \etal~1999, \aj, 118, 1766
\bibitem[Richmond, Treffers \& Filippenko (1993)]{rich93}Richmond, M. W., Treffers, R. R., \& Filippenko, A. V. 1993, PASP, 105, 1164
\bibitem[Richmond \etal~(1995)]{rich95}Richmond, M. W., \etal~1995, \aj, 109, 2121 
\bibitem[Riess (2000)]{ries00}Riess, A. G. 2000, in Cosmic Flows Workshop, eds. S. Courteau \& J. Willick (San Francisco: ASP), p. 80 
\bibitem[Riess \etal~(1998b)]{ries98b}Riess, A. G., Nugent, P., Filippenko, A. V., Kirshner, R. P., \& Perlmutter, S. 1998b, \apj, 504, 935 
\bibitem[Riess, Press, \& Kirshner (1995)]{ries95}Riess, A. G., Press, W. H., \& Kirshner, R. P. 1995, \apj, 445, L91 
\bibitem[Riess, Press, \& Kirshner (1996)]{ries96}---------. 1996, \apj, 473, 88 
\bibitem[Riess \etal~(1998a)]{ries98a}Riess, A. G., \etal~1998a, \aj, 116, 1009 

\bibitem[Riess \etal~(1999a)]{riesetal99a}Riess, A. G., \etal~1999a, \aj, 117, 707  
\bibitem[Riess \etal~(1999b)]{riesetal99b}---------. 1999b, \aj, 118, 2675 
\bibitem[Ruiz-Lapuente \etal~(1992)]{ruiz92}Ruiz-Lapuente, P., Cappellaro, E., Turatto, M., Gouiffes, C., Danziger, I. J., Della Valle, M., \& Lucy, L. B. 1992, \apj, 387, L33
\bibitem[Ruiz-Lapuente \etal~(1993)]{ruiz93}Ruiz-Lapuente, P., \etal~1993, \nat, 365, 728
\bibitem[Saha \etal~(2000)]{saha00}Saha, A., Labhardt, L., \& Prosser, C. 2000, \pasp, 112, 163
\bibitem[Saha \etal~(1995)]{saha95}Saha, A., Sandage, A., Labhardt, L., Schwengeler, H., Tammann, G. A., Panagia, N., \& Macchetto, F. D. 1995, \apj, 438, 8
\bibitem[Saio \& Nomoto (1985)]{saio85}Saio, H., \& Nomoto, K. 1985, \aap, 150, L21
\bibitem[Saio \& Nomoto (1998)]{saio98}---------. 1998, \apj, 500, 388
\bibitem[Savage \& Mathis (1979)]{sava79}Savage, B. D., \& Mathis, J. S. 1979, \araa, 17, 73
\bibitem[Schlegel, Finkbeiner, \& Davis (1998)]{schl98}Schlegel, D. J., Finkbeiner, D. P., \& Davis, M. 1998, \apj, 500, 525
\bibitem[Schmidt \etal~(1993)]{schm93}Schmidt, B. P., \etal~1993, \aj, 105, 2236
\bibitem[Stetson (1987)]{stet87}Stetson, P. B. 1987, PASP, 99, 191
\bibitem[Suntzeff \etal~(1999)]{sunt99}Suntzeff, N., \etal~1999, \aj, 117, 1175
\bibitem[Tonry \etal~(2000)]{tonr00}Tonry, J. L., Blakeslee, J. P., Ajhar, E. A., \& Dressler, A. 2000, \apj, 530, 625
\bibitem[Treffers \etal~(1997)]{tref97}Treffers, R. R., Peng, C. Y., Filippenko, A. V., \& Richmond, M. W. 1997, \iaucirc~6627
\bibitem[Tripp (1998)]{trip98}Tripp, R. 1998, \aap, 331, 815
\bibitem[Turatto \etal~(1998)]{tura98}Turatto, M., Piemonte, A., Benetti, S., Cappellaro, E., Mazzali, P. A., Danziger, I. J., \& Patat, F. 1998, \aj, 116, 2431
\bibitem[Turatto \etal~(1996)]{tura96}Turatto, M., \etal~1996, MNRAS, 283, 1 (T96)
\bibitem[Vaughan \etal~(1995)]{vaug95}Vaughan, T. E., Branch, D., Miller, D. L., \& Perlmutter S. 1995, \apj, 439, 558
\bibitem[Vettolani \& Baiesi Pillastrini (1987)]{vett87}Vettolani, G., \& Baiesi Pillastrini, G. C. 1987, \aap, 175, 9
\bibitem[Wade \& Horne (1988)]{wade88}Wade, R. A., \& Horne, K. 1988, \apj, 324, 411
\bibitem[Webbink (1984)]{webb84}Webbink, R. F. 1984, \apj, 277, 355
\bibitem[Wells \etal~(1994)]{well94}Wells, L. A., \etal~1994, \aj, 108, 2233
\bibitem[Wheeler \& Harkness (1990)]{whel90}Wheeler, J. C., \& Harkness, R. P. 1990, Rep. Prog. Phys., 53, 1467
\bibitem[Woosley \& Weaver (1986)]{woos86}Woosley, S. E., \& Weaver, T. A. 1986, \araa, 24, 205
\bibitem[Woosley \& Weaver (1994)]{woos94}---------. 1994, \apj, 423, 371
\bibitem[Zehavi \etal~(1998)]{zeha98}Zehavi, I., Riess, A. G., Kirshner, R. P., \& Dekel, A. 1998, \apj, 503, 483

\end{thebibliography}
\end{document}